\documentclass[aps,prl,twocolumn,amsfonts,showpacs,longbibliography,superscriptaddress]{revtex4-2}
\usepackage{verbatim,epsfig,amsmath,amssymb,bm,epsf,graphicx,psfrag,bbold,amsthm,amsfonts}
\usepackage[export]{adjustbox}
\usepackage[toc,page,titletoc]{appendix}
\usepackage[bottom]{footmisc}
\usepackage[hypertexnames=false]{hyperref}
\usepackage[capitalize]{cleveref} 
\hypersetup{
	colorlinks=true,
	linkcolor=black,
	citecolor=blue,
	filecolor=black,
	urlcolor=blue,
}
\usepackage{enumitem}
\usepackage{framed}
\usepackage{mathrsfs}
\usepackage{esint}
\setlist{nosep}
\usepackage{color}
\usepackage[utf8]{inputenc}
\usepackage{float}
\usepackage{natbib}
\usepackage{tikz-cd}
\usepackage{leftidx}
\usepackage[normalem]{ulem}
\usepackage[caption=false]{subfig}
\usepackage{notes2bib}
\usepackage{physics}
\usepackage{placeins}

\newcommand{\moy}[1]{\langle #1 \rangle}

\newcommand{\Ham}{\mathrm{H}}
\newcommand{\cM}{\mathcal{M}}
\newcommand{\bZ}{ \mathbb{Z}}

\theoremstyle{plain}

\usepackage{cancel}
\theoremstyle{definition}

\theoremstyle{remark}

\newcommand{\sI}{\mathscr{I}}
\newcommand{\sC}{\mathscr{C}}
\renewcommand{\arraystretch}{1.15}
\makeatletter
\renewcommand{\l@subsubsection}[2]{}
\makeatother

\begin{document}
\title{\large Topological Phenomena Protected by Diabolical Textures}
	\author{Sayantan Mandal}
    \email{sayantan.mandal@utdallas.edu}
\affiliation{Harish-Chandra Research Institute (HRI), Prayagraj (Allahabad) 211019, India}
\affiliation{University of Texas at Dallas, Richardson, Texas 75080, USA}

	\author{Neelima Pulletikurty}
    \email{422ph5007@nitrkl.ac.in}
  \affiliation{Harish-Chandra Research Institute (HRI), Prayagraj (Allahabad) 211019, India}
    \affiliation{National Institute of Technology (NIT), Rourkela 769008, India}

	\author{Abhishodh Prakash}
    \email{abhishodhprakash@hri.res.in}
    \altaffiliation{(he/him/his)}
\affiliation{Harish-Chandra Research Institute (HRI), Prayagraj (Allahabad) 211019, India}
\affiliation{Homi Bhabha National Institute (HBNI),  Mumbai 400094, India}
	
\begin{abstract}
We present a new class of topological phenomena in inhomogeneous systems arising from the adiabatic spatial embedding of parametrized families of quantum states such as charge pumps and their generalizations. We demonstrate that each topologically distinct class of these “diabolical textures” gives rise to distinct gapped states that are separated by “trap-scaling” critical points, defect transitions, or intervening phases. Their phase diagrams exhibit versions of  ``unnecessary criticality'' and ``multiversality,''  within a noninteracting setting. We demonstrate this in a one-dimensional Thouless pump texture trapping a charge and a two-dimensional class D texture trapping a Kitaev chain, including an analysis of their stability under weak symmetry-preserving perturbations and interactions. For systems in arbitrary spatial dimensions and global symmetries, we present a framework to systematically classify diabolical textures using Kitaev's $\Omega$ spectrum conjecture.
\end{abstract}

\maketitle
\noindent The past few decades have seen the emergence of a broad class of quantum phases distinguished not by spontaneous symmetry breaking but by topology~\cite{Haldane_Nobel_RevModPhys.89.040502,Thouless_Nobel_RevModPhys.89.040501}. The simplest among these, invertible
phases, possess unique ground states on closed manifolds with robust properties such as protected anomalous boundary modes and quantized responses to gauge
fields. A landmark achievement
is the complete classification of free-fermion topological
phases~\cite{SchnyderRyu,Ryu,kitaevclassification} together with a conjectural framework for
free and interacting systems~\cite{kitaev1,kitaev2,kitaev3,kubota2025stablehomotopytheoryinvertible,Xiong_2018,Gaiotto_2019}. A closely related development is the study of \emph{topological families} of gapped states~\cite{kitaev2019,Wenetal_topologicalfamilies,HsinKapustinThorngren_PhysRevB.102.245113,Shiozaki_SPTPUmp_PhysRevB.106.125108,JonesThorngrenAP_SPTPump,AP_UC_Pump,AP_ThoulessPhasediagrams}. Familiar examples
include the Thouless pump and Berry's quantum spin in a magnetic field, where a loop or sphere in parameter space
cannot be contracted without encountering a gapless obstruction. The singularities enclosed by such surfaces are called \emph{diabolical points}~\cite{HsinKapustinThorngren_PhysRevB.102.245113,manjunath2026searchdiabolicalcriticalpoints}. Phase diagrams containing topological families were recently shown to have topological textures with diabolical points at the core~\cite{AP_ThoulessPhasediagrams}.  

In this work, we show that spatially embedding such topological families produces a distinct class of topological phenomena. By varying Hamiltonian parameters adiabatically along space, one imprints a spatial \emph{diabolical texture} on the many-body ground state. Different topological classes of textures label distinct gapped regimes. Our first example is
a one-dimensional Class A free-fermion chain in which the
couplings of the Rice–Mele model~\cite{RMmodel} are varied slowly along
the chain so as to realize a spatial Thouless pump~\cite{AsagaFukui2021}. The non-trivial texture introduces an additional charged mode delocalized over a sub-extensive region on the chain. Removing this texture and charge necessarily closes the bulk excitation gap at
a stable critical point with an unconventional finite-size
scaling characteristic of trap-scaling criticality~\cite{CampostriniVicari2009PRL_TSS,CampostriniVicari2010PRA_QTSS,CeccarelliTorreroVicari2013PRB_CriticalParamsTSS}. We further show that sharpening the spatial texture
causes the trap-scaling critical line to terminate abruptly, where the charged mode is confined to a defect or boundary, and  
expelled by a single-particle level crossing. This realizes
a form of \emph{unnecessary criticality}~\cite{BiSenthil,AP_UC_Multiversality,AP_UC_Pump}. At finite chemical potential, critical lines broaden into metallic lobes whose phase boundaries support inequivalent trap-scaling critical theories along different segments and realize a form of \emph{multiversality}~\cite{BiSenthil,AP_UC_Multiversality}. To our knowledge, these are the first free-fermion realizations of unnecessary criticality and multiversality, albeit in an unconventional trap-scaling setting. We also study a two-dimensional class D example, whose texture traps a Kitaev chain with an analogous rich phase diagram. We analyze the stability of these structures under weak symmetric perturbations. Finally, for arbitrary dimensions, symmetry, and interactions, we give a general classification of diabolical textures using Kitaev’s $\Omega$-spectrum
conjecture~~\cite{kitaev1,kitaev2,kitaev3,kubota2025stablehomotopytheoryinvertible,Xiong_2018,Gaiotto_2019}.

\medskip
\begin{figure*}[!ht]
    \centering
    \includegraphics[height=.31\linewidth]{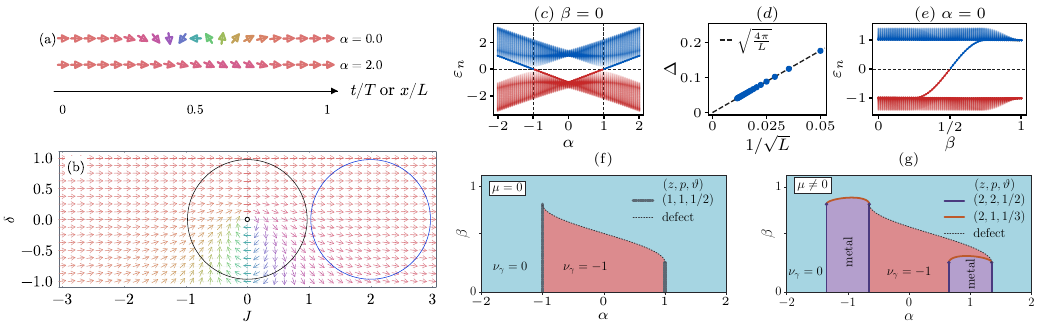}
    \caption{(a,b) The Thouless charge pump can be interpreted as a non-trivial temporal texture generated by dynamically encircling the diabolical point of the Rice-Mele model. The texture and the corresponding charge pump are trivial when the diabolical point is not enclosed. (c) These textures can also be embedded in space, where a non-trivial texture causes the ground state to bind an extra charge. As the non-trivial texture is eliminated by tuning $\alpha$ in \cref{eq:Thouless_parameters_space}, the bulk excitation gap closes as the charge is expelled at the trap-scaling critical points $\alpha = \pm 1$, with the neutral gap exhibiting trap-critical finite-size scaling $\Delta\sim1/\sqrt{L}$, shown in (d) for $\alpha=1$, $\mu=\beta=0$. (e) By tuning $\beta$ in \cref{eq:Thouless_parameters_space_beta}, the texture and trapped charge can be squeezed into the boundary (for open boundary conditions) or a defect (for periodic boundary conditions), and expelled through a level crossing. Spectra in (c,e) are for an open chain with $L=200$. (f,g) Phase diagrams for the $\beta$ deformation. At $\mu=0$ (f), the trap-critical lines terminate; for $\mu\ne0$ (g), they broaden into metallic lobes whose phase boundaries exhibit multiversality where different segments host distinct trap-scaling criticality. Regions with different $\nu_\gamma$ are separated by the terminating critical lines at $\mu=0$, by the metallic lobes at $\mu\ne0$, or by a defect crossing.}
    \label{fig:texture}
\end{figure*}

\noindent \textbf{\emph{Thouless pump as a temporal texture:}} The Thouless charge pump is a dynamical phenomenon characterized by quantized charge transport as the system parameters are adiabatically varied in a cycle~\cite{Thouless1983paper}. A representative model is the Rice-Mele chain~\cite{RMmodel},
\begin{equation}
    \Ham_{\text{RM}} = \sum_{x=1}^L \left(t_x(\delta)(c^\dagger_x c_{x+1} + h.c.) + (-1)^x J c^\dagger_x c_x\right),\label{eq:H_Rice_Mele}
\end{equation}
where $t_x(\delta) = \left(1-(-1)^x \delta \right)/2$. At half filling, the ground state of \cref{eq:H_Rice_Mele} is a trivial band insulator for all parameter values except at the origin $J=\delta=0$ when the single-particle bands touch. A Thouless charge pump results if $\delta$ and $J$ are adiabatically varied in time so as to enclose this \emph{diabolical point}. For instance, we may take
\begin{equation}
    \delta(t) =\cos\left( \frac{2 \pi t}{T} \right),~J(t) =\alpha + \sin\left( \frac{2 \pi t}{T} \right), \label{eq:Thouless_parameters_time}
\end{equation}
with $|\alpha|<1$ and $T\gg 1$. The topological nature of this dynamical process can be visualized as a temporal texture following Ref.~\cite{AP_ThoulessPhasediagrams}. For each instant in time $t$ we compute the Berry-Bloch connection $A_k = i \innerproduct{\varepsilon^-_k}{\partial_k|\varepsilon^-_k}$ and Berry phase $\gamma=\oint_{\text{BZ}} dk A_k$~\cite{KaneTopologicalBandTheoryReview} for the filled band $\ket{\varepsilon^-_k}$. Plotting $\gamma(t)$ reveals this texture. As shown in \cref{fig:texture}(a), for $|\alpha|<1$, the texture is non-trivial and exhibits a winding, whereas for $|\alpha|>1$, it is trivial. It is also illuminating to plot $\gamma$ over the $(J,\delta)$ phase diagram, as shown in \cref{fig:texture}(b) which reveals a vortex-like texture~\cite{AP_ThoulessPhasediagrams} with the diabolical point at the core. As the Hamiltonian parameters  trace a closed loop in the phase diagram, the system develops a temporal texture whose topology depends on whether or not the loop encloses the diabolical point. We will refer to these as `diabolical textures'.  $\alpha = \pm 1$ represent the critical points that separate the trivial and non-trivial pump. The Hamiltonian strikes the gapless point $\delta = J = 0$ along the cycle at $t^* = 3T/4$ and $T/4$ respectively, and the adiabatic limit is lost for any finite $T$.
Defining $s=t/T$,  Berry connection component 
$A_s=i\langle\varepsilon^-_k|\partial_s|\varepsilon^-_k\rangle$ and denoting $e$ the charge of the electron, the
electric polarization and the first Chern number of the pump are
related by~\cite{KingSmithVanderbilt1993,Thouless1983paper,KaneTopologicalBandTheoryReview}
\begin{align}
 P(s)&=e\frac{\gamma(s)}{2\pi}\pmod e,\qquad
 F_{sk}=\partial_sA_k-\partial_kA_s,\nonumber\\
 C&=\frac{1}{2\pi}\int_0^1ds\oint_{\mathrm{BZ}}dk\,F_{sk}=\frac{\gamma(1)-\gamma(0)}{2\pi}\in\bZ,
 \label{eq:polarization_chern}
\end{align}
With electric current
$j=\partial_tP$, the charge transported during one temporal cycle is $Q_{\mathrm{pump}}=eC$.

\medskip

\noindent \textbf{\emph{Thouless pump in a spatial texture}:}  We now modify the system so that the variation of $\delta$ and $J$ occurs in space rather than time, by replacing $t \rightarrow x$ in \cref{eq:Thouless_parameters_time}:
\begin{equation}
   \begin{pmatrix}
   	\delta(x) \\
   	 J(x)
   \end{pmatrix}  =\begin{pmatrix}
   	\cos(\zeta(x))\\
   	\alpha + \sin\left(\zeta(x)\right)
   \end{pmatrix}, \quad \zeta(x) =\left(\frac{2 \pi x}{L} \right) \label{eq:Thouless_parameters_space}
\end{equation}
By construction,
$\partial_x\delta,\partial_xJ=\mathcal{O}(L^{-1})$.  At each position we
define the \emph{frozen} Bloch Hamiltonian
$h(k;x)=h_{\mathrm{RM}}[k;\delta(x),J(x)]$, with local interband gap
$\Delta_{\mathrm{loc}}(x)$, correlation length $\xi(x)$ and Berry
phase $\gamma(x)$. $P(x)=e\frac{\gamma(x)}{2\pi}$ can be interpreted as the local
 polarization of the inhomogeneous ground state when $\xi(x)\ll L_{\mathrm{tex}}(x)$ where $L_{\mathrm{tex}}$ is the coupling-variation
scale~\cite{KingSmithVanderbilt1993,TeoKane2010TopologicalDefects,supp}.  For
\cref{eq:Thouless_parameters_space} this holds at fixed
$|\alpha|\ne1$ as $L\to\infty$, but fails at
$|\alpha|=1$.  We can define the winding number,
$\nu_\gamma= \frac{1}{2\pi}\int dx \partial_x \gamma(x)$ which is nonzero for $|\alpha|<1$ and vanishes for
$|\alpha|>1$.  How should we understand the two gapped regions distinguished by $\nu_\gamma$, and what
happens at $\alpha^*=\pm1$?

The temporal and spatial effects are two
components of the same polarization
response~\cite{GoldstoneWilczek1981,KingSmithVanderbilt1993,AsagaFukui2021}.  Denote $\delta q(x)$ as the excess  charge density relative to
the locally homogeneous half-filled background,
\begin{multline}
 \delta q=-\partial_xP
 =-e\frac{\partial_x\gamma}{2\pi}
 +\mathcal{O}(\partial_x^2) \implies  Q_{\mathrm{tex}}=-e\nu_\gamma .
 \label{eq:texture_charge_chern}
\end{multline}
Here $\nu_\gamma=C$ because $x/L$ and $s$ parametrize the same oriented loop.
For the orientation in
\cref{eq:Thouless_parameters_space}, $C=-1$ when $|\alpha|<1$, so the
static texture binds electric charge $Q_{\mathrm{tex}}=e$~\cite{HsinKapustinThorngren_PhysRevB.102.245113,Shiozaki_SPTPUmp_PhysRevB.106.125108}. To see this, we numerically compute the flow of the single-particle energy levels as $\alpha$ is tuned. As shown in \cref{fig:texture}(c), $|\alpha|<1$ contains an additional occupied mode below the Fermi level, which is expelled
into the conduction band for $|\alpha|>1$ when the diabolical texture is removed. Charges trapped by spatial textures bear a family resemblance to a similar phenomenon in quantum Hall ferromagnets~\cite{Sondhi_QH_Skyrmion_PhysRevB.47.16419,Else_TopologicalGoldstone_PhysRevB.104.115129,Else_Manjunath_TopologicalGoldstone_PhysRevB.111.125151}. For our system, there exist no spontaneously broken symmetries or order parameters. Thus, the two regions differ by one unit of electric charge, an $\mathcal O(1)$ difference invisible in the bulk charge density. Nevertheless, charge expulsion closes the bulk excitation gap [\cref{fig:texture}(c)]. However, as shown in \cref{fig:texture}(d) the finite-size neutral gap scaling takes the unusual form $\Delta \sim L^{-1/2}$ compared to conformal ($\Delta \sim L^{-1}$) or Lifshitz ($\Delta \sim L^{-2}$)  criticality.

\medskip 

\noindent \textbf{\emph{Trap-scaling criticality}:}  To understand the nature and stability of the charge expelling critical points, we use the local spectral gap, $\Delta_{\mathrm{loc}}(x)$ $\propto  \sqrt{\delta^2(x) + J^2(x)}$ corresponding to the approximate homogeneous system at each position which gives the energy scale of excitations near $x$.  At the critical points $\alpha^* = \pm 1$, the system becomes locally gapless near $x^* = 3L/4$ and $L/4$ respectively. We therefore expect the low-energy critical modes to be trapped in the
vicinity of  $x^*$, which we will refer to as the \emph{trapping node} following Refs~\cite{CampostriniVicari2009PRL_TSS,CampostriniVicari2010PRA_QTSS,CeccarelliTorreroVicari2013PRB_CriticalParamsTSS}. We can write down a continuum theory that reproduces the low-energy physics near  $\alpha \approx \alpha^*$ and $x \approx x^*$:
\begin{align}
 \Ham_{\text{RM}} &\approx \int dx~ \psi^\dagger(x) h(x) \psi(x), \nonumber\\
 h(x) &=  \left( i \sigma_2   \partial_x 
 +\text{sgn}(\alpha^*) \frac{(x -x^*)}{\ell^2} \sigma_1  - m  \sigma_3 \right),
 \label{eq:Dirac_trapnode}
\end{align} 
 where $\ell = \sqrt{L/(2\pi)}$, $m  = \alpha -\alpha^*$ and $\psi(x)$ is a two-component spinor. Formally, \cref{eq:Dirac_trapnode} is equivalent to Landau-level problem for a two-dimensional relativistic Dirac fermion in Landau gauge, with $\ell$ playing the role of the magnetic length, $x^*$ the guiding center and $m$ the mass~\cite{supp}. Thus, the single-particle spectrum is 
\begin{equation}
     \varepsilon_0 =\text{sgn}(\alpha^*) m,~\varepsilon_n^\pm = \pm\sqrt{m^2+\frac{2n}{\ell^2}},~n=1,2,\ldots. \nonumber 
\end{equation}
Away from criticality ($m \neq 0$), the system has a finite neutral excitation gap $\Delta \sim |m|$. At criticality ($m=0$), the nonzero trap levels and gap scale as $\Delta \sim \ell^{-1} \sim L^{-1/2}$ in agreement with numerics. The low-energy modes are supported over a length $\ell \sim \sqrt{L}$ around the trapping node. This is the hallmark of \emph{trap-scaling universality}~\cite{CampostriniVicari2009PRL_TSS,CampostriniVicari2010PRA_QTSS,CeccarelliTorreroVicari2013PRB_CriticalParamsTSS}. It is characterized by a trap-scaling critical exponent $\vartheta$~\cite{CeccarelliTorreroVicari2013PRB_CriticalParamsTSS}. More generally, suppose the coefficient of a relevant operator $\hat M$ of scaling dimension $\chi$ vanishes near the trapping node as $|x-x^*|^p/L^p$, where $p$ is the order of the zero. At a critical point in $d$ spatial dimensions with dynamical exponent $z$, we have  $\vartheta=p/(p+d+z-\chi)$ such that $\ell\sim L^\vartheta$ and $\Delta\sim L^{-z\vartheta}$~\cite{endmatter}. This introduces corrections to finite-size scaling of local observables  and von Neumann entanglement entropy~\cite{Cardy_ScalingRG_StatPhys_1996,Cardy_Calabrese_2004,CampostriniVicari2010Entanglement}.
For the noninteracting Rice--Mele chain, we have $\vartheta = 1/2$. 

\medskip

\noindent\textbf{\emph{Stability to perturbations}:}  At fixed chemical potential $\mu=0$, the two Dirac masses in \cref{eq:Dirac_trapnode} saturate the relevant perturbations~\cite{AP_ThoulessPhasediagrams}; weak single-particle perturbations may shift $\alpha^*$ and $x^*$ but cannot destroy the critical line. Short-range interactions are marginal and renormalize the operator scaling dimensions giving $\vartheta=(3-K)^{-1}$, where $K$ is the so-called Luttinger parameter. Throughout an open interval about the noninteracting limit $K=1$, no additional symmetry-allowed operator is relevant and the phase diagram remains unchanged~\cite{Giamarchi,HALDANE_Bethe_LL_1981153,AP_SEC_PhysRevB.108.245135,supp}.

\medskip 

\noindent \textbf{\emph{Unnecessary criticality}:} 
 The robustness of the two textured regimes and the transition between them raises the natural question: are they genuinely distinct phases of matter? The only essential symmetry is charge conservation, placing the model in class A, which has no non-trivial strong topological phases in 1d~\cite{kitaevclassification}. Moreover, our system lacks translation invariance, ruling out weak topology~\cite{HasanKane_RevModPhys.82.3045}, and the continuum analysis shows that adding extra filled or empty bands does not affect stability, excluding fragile topology~\cite{PoWatanabeVishwanath_FragileTopology_PhysRevLett.121.126402}. We will argue that the two regions are, strictly speaking, the same phase of matter. We modify the spatial texture in \cref{eq:Thouless_parameters_space} to
 \begin{equation}
   \zeta(x) \rightarrow \zeta_\beta(x) =\left(\frac{2 (1-\beta)\pi x}{L} \right) \label{eq:Thouless_parameters_space_beta}   
\end{equation}
So far, we have not specified the boundary conditions. Note that \cref{eq:Thouless_parameters_space} is compatible with both periodic (PBC) and open boundary conditions (OBC), since $\delta(x+L) = \delta(x),~J(x+L) = J(x)$.  The interpretation of \cref{eq:Thouless_parameters_space_beta} depends on this choice. For OBC, this retains the slowly varying texture while changing the endpoint values, thereby ``unwinding'' the texture as we tune from 0 to 1. For PBC, it introduces a defect near the link connecting $x=L$ and $x=1$ across which the parameters jump rapidly and the local gradient expansion fails.

Regardless of the boundary conditions, for small $\beta$ the distinction between the regions $|\alpha|<1$ and $|\alpha|>1$ as well as the critical points $\alpha^* =\pm1$ remain stable. At $\alpha = \alpha^*$, the effect of $\beta$ is to shift the trapping node to $x^* = 3L/(4(1-\beta))$ and $x^* = L/(4(1-\beta))$, respectively, towards the boundary (OBC) or the defect (PBC). For $\beta^* = 1/4~(3/4)$ for $\alpha^* =1~(-1)$, they reach $x^* = L$ and are expelled from the system or absorbed by the defect.  The critical line abruptly terminates  [\cref{fig:texture}(f)], realizing \emph{unnecessary criticality}~\cite{BiSenthil,Xu_UC_PhysRevB.101.035118,verresen2021quotientsymmetryprotectedtopological,AP_UC_Multiversality,AP_UC_Pump,AP_J1J2,Senthil_UC_Dirac_SciPostPhysCore.8.1.024,Senthil_UC_Kagome_SciPostPhys.20.3.082,APJones_ClassicalDQC_PhysRevLett.134.097103}. To our knowledge, this is the first in a free-fermion setting~\cite{BiSenthil,AP_UC_Multiversality}.

	Within the non-trivial region in $|\alpha|<1$, $\beta$ squeezes the delocalized charged mode into the boundary (OBC) or the defect (PBC), where it is eventually expelled through a single level crossing at the Fermi level without closing the bulk excitation gap. This crossing exchanges many-body ground states with different $\mathrm{U}(1)$ charges.  They cannot hybridize, so weak interactions can shift but not avoid the crossing; together with the stable trap-scaling line, this preserves the terminating structure while renormalizing its location~\cite{AP_UC_Pump}. For the open termination in \cref{fig:texture}(f), the crossing occurs at $\alpha_{\rm def}(\beta)=\sin(2\pi\beta)$ and joins the terminal points at $\alpha=\pm1$~\cite{supp}. Thus, the $\nu_{\gamma}=0,-1$ regions can be connected without encountering any thermodynamic singularities. If one adopts the somewhat unconventional perspective~\cite{AP_UC_Pump,Khalaf_BOTP_PhysRevResearch.3.013239} of treating the boundary/ defect crossing on the same footing with bulk criticality, the two regimes are sharply distinct.

\medskip
\noindent \textbf{\emph{Finite chemical potential}:} For
$0<\mu<1$, the Dirac trap-critical lines broaden into 
metallic phases [\cref{fig:texture}(g)] when the conduction-band minimum dips below the chemical potential, $\min_k\varepsilon_+(k,x)<\mu$. In the $(\alpha,\beta)$ phase diagram, this produces two metallic lobes; together with the open-boundary crossing locus $\alpha_{\rm def}(\beta)=\sin(2\pi\beta)+\mu$, $1/4<\beta<3/4$, they separate regions with distinct textures. For small $\beta$, the lobes occupy $\alpha^{*}-\mu<\alpha<\alpha^*+\mu$ and contain an extensive droplet centered on $x^{*}$, of width $w=\frac{L}{\pi(1-\beta)}\arccos\left(\frac{1+\alpha^2-\mu^2}{2|\alpha|}\right)$ before either droplet interface reaches the end of the chain. Increasing $\beta$ pushes this droplet to the end (OBC) or defect (PBC), where the lobe closes along a curved phase boundary (cap) [\cref{fig:texture}(g)]. With unit lattice spacing, expanding the band edge gives
 \begin{equation}
\varepsilon_+-\mu\simeq a (k-\pi/2)^2+
\begin{cases}
c_2[(x-x^*)/L]^2,&\text{lobe sides},\\
c_1(L-x)/L,&\text{lobe caps},
\end{cases} \nonumber
 \end{equation}
where $a,c_1,c_2>0$. The quadratic dispersion gives $z=2$  so $\vartheta=p/(p+2)$. The straight phase boundaries have $(z,p,\vartheta)=(2,2,1/2)$ and trap-level spacing $\Delta\sim L^{-1}$, whereas the curved phase boundaries have $(2,1,1/3)$ and $\Delta\sim L^{-2/3}$~\cite{supp}. Distinct universality classes along a single connected metal--insulator phase boundary realize \emph{multiversality}~\cite{BiSenthil,AP_UC_Multiversality}. To our knowledge, this is also the first in a free-fermion setting.

\medskip
\noindent \textbf{\emph{Generalization}:} Diabolical textures reveal a connection between families of
states and lower-dimensional phases as illustrated by the Rice--Mele texture: as the spatial coordinate is traversed, the local
Hamiltonian traces a closed path of one-dimensional class A Hamiltonians. This loop is classified by its trapped $\mathrm U(1)$ charge, a
zero-dimensional phase.  This generalizes: 
a loop texture in a $d$-dimensional system traps a $(d-1)$-dimensional
invertible phase~\footnote{An invertible phase is a gapped phase with an inverse under stacking: stacking it with its inverse can be deformed to a trivial product state without closing the gap or breaking the specified symmetry. Such phases have no intrinsic topological order and a unique ground state in the absence of boundaries; examples include symmetry-protected topological phases and integer quantum Hall phases~\cite{kitaev1,kitaev2,Gaiotto_2019,kubota2025stablehomotopytheoryinvertible}.}. Restricting a texture $\varrho_k$ to a spherical
$k$-cycle on spatial manifold $\cM_d$  traps a
$(d-k)$-dimensional invertible phase. This follows from Kitaev's conjecture that stabilized spaces of gapped
invertible Hamiltonians, with fixed symmetry and statistics and varying spatial
dimension, form an $\Omega$-spectrum~\cite{kitaev1,kitaev2,kitaev3,kubota2025stablehomotopytheoryinvertible}.
 The full classification is given by the generalized-cohomology group $[\varrho] \in h_G^d(\cM_d)$~\cite{endmatter,supp}. $\varrho_0$ is nothing but the $d$-dimensional phase to which each
connected component of $\cM_d$ belongs. Thus, diabolical textures are additional
decorations on top of the underlying phase.

\begin{figure}[H]
 \centering
 \includegraphics[width=\columnwidth]{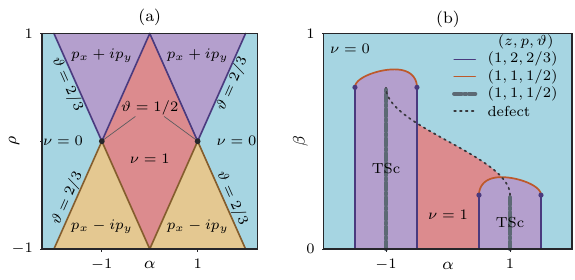}
 \caption{Phase diagrams of the textured class D Hamiltonian in
 \cref{eq:classD_h0,eq:classD_MD_operator,eq:classD_texture3} with (a) $\beta=0$ and
 (b) the $\rho=0$ and $\rho=1/2$ phase boundaries superposed, for open boundary. The $\nu=0,1$ regions are separated by terminating trap-scaling transitions ($\rho=0$), intervening topological superconductor (TSc) phases ($\rho \neq 0$) or defect transitions. TSc phase boundaries  exhibit trap-scaling multiversality.}
 \label{fig:classD_phase}
\end{figure}

\medskip
\noindent {\clubpenalty=10000
\textbf{\emph{A two-dimensional example}:} As an illustration of this generalization, we consider a loop family of
two-dimensional class D systems whose spatial texture traps Kitaev's Majorana chain~\cite{KitaevMajoranaChain,kitaevclassification,Ryu}.
We obtain such a family by the suspension construction: stack Majorana chains and deform the interlayer pairing in the two inequivalent
ways~\cite{AP_ThoulessPhasediagrams,Wenetal_topologicalfamilies,endmatter,supp}. The Hamiltonian, with
$a_{x,y}$ and $b_{x,y}$ for the two Majoranas, with $x$
along the chains and $y$ labeling an even stack of layers, is
\begin{multline}
 \Ham_0=\frac{i}{4}\sum_{x,y}\Big[
 (1+m)a_{x,y}b_{x,y}+(1-m)b_{x,y}a_{x+1,y}\\[-2pt]
+\big(1+(-1)^y\delta\big)
 \big(a_{x,y}a_{x,y+1}-b_{x,y}b_{x,y+1}\big)\Big].
 \label{eq:classD_h0}
\end{multline}
These loop families carry a $\mathbb Z_2$ Chern--Simons
invariant over $X=T^2_{\rm BZ}\times S^1_\lambda$, where $\lambda$
parametrizes the loop.  
\begin{equation*}
 \nu=\frac{\epsilon_{\mu\nu\sigma}}{4\pi^2}\int_Xd^3\xi\,\operatorname{Tr}\!\Bigl(A^\mu\partial_\nu A^\sigma-\frac{2i}{3}A^\mu A^\nu A^\sigma\Bigr)\pmod2.
\end{equation*}
Here $\vec{\xi}=(k_x,k_y,\lambda)$,
$\partial_\mu=\partial/\partial\xi^\mu$, and
$A^\mu_{mn}=i\langle u_m|\partial_\mu |u_n\rangle$ is the non-Abelian Berry connection for occupied Bogoliubov--de Gennes (BdG) states
$|u_n\rangle$~\cite{TeoKane2010TopologicalDefects,supp}. The gap of $\Ham_0$ closes only at $m=\delta=0$. A loop around this diabolical point has $\nu=1$. Embedding this loop along the $y$ direction produces a diabolical texture.
We measure $y,L_y$ in two-layer unit cells.
\begin{equation}
	m(y) = \alpha + \sin \left(\frac{2\pi y}{L_y}\right),\quad \delta(y) = \cos \left(\frac{2\pi y}{L_y}\right)
	\label{eq:classD_texture3}
\end{equation}
The texture has $\nu=1$ for $|\alpha|<1$ and $\nu=0$ for $|\alpha|>1$. Tuning $\alpha$ through $\alpha^*=\pm1$  produces a critical theory of two relativistic Majoranas~\cite{ReadGreen2000,Ryu} in a linear trap,
\begin{align}
 \Ham_0&\approx\frac12\int d^2r\,
 \Psi^\dagger(\vec{r})h(\vec{r})\Psi(\vec{r}),\label{eq:classD_MD0_trap}\\
 h&=-\frac{i}{2}\left(\Gamma_1\partial_x+\Gamma_2\partial_y\right)+(\alpha-\alpha^*)\Gamma_3
 +s_*\frac{(y-y^*)}{\ell_y^2}\Gamma_4,
 \nonumber
\end{align}
where $\Psi(\vec{r})=\bigl[\Psi^\dagger(\vec{r})\bigr]^T$ is a four-component self-conjugate BdG field,
$s_*=\alpha^*=\pm1$, $y^*/L_y=(3/4,1/4)$ for
$\alpha^*= \pm 1$, $\ell_y=\sqrt{L_y/(2\pi)}$ and $\{\Gamma_a,\Gamma_b\}=2\delta_{ab}$~\cite{supp}.  
At $\alpha = \alpha^*$, the subextensive critical droplet is of size $\sim L_x \ell_y$, i.e.,
$\vartheta=1/2$~\cite{CampostriniVicari2009PRL_TSS,CampostriniVicari2010PRA_QTSS,endmatter}.
This two-Majorana trap-scaling criticality is not generic in class D.  Fermion parity
allows the relevant perturbation $\delta h=  - i \rho \Gamma_1 \Gamma_2$ at field theory level. On the lattice, this is introduced by  
\begin{equation}
 \begin{aligned}
  \Ham_{\rho}&=\frac{i\rho}{4}\sum_{x,y}
 \left(a_{x,y}b_{x,y+1}-b_{x,y}a_{x,y+1}\right)
 \end{aligned}
 \label{eq:classD_MD_operator}
\end{equation}

For $0<|\rho|<1$, this perturbation produces an intervening
$p_x\pm i p_y$ topological superconductor (TSc) phase
[\cref{fig:classD_phase}].
For $1-|\rho|<|\alpha|<1+|\rho|$, the TSc is an extensive strip of width $ w=\frac{L_y}{\pi}\arccos\!\left(
 \frac{1+\alpha^2-\rho^2}{2|\alpha|}\right)$ along $y$, with a chiral
Majorana mode on each of its two interfaces
~\cite{ReadGreen2000,TeoKane2010TopologicalDefects}. Projecting the lattice Hamiltonian onto the light Majorana sector gives the effective trap theory describing the transition out of the TSc. For $\rho>0$, near $\alpha_c = 1+\rho$, this takes the form
\begin{align}
 \Ham&\approx\frac12\int d^2r\,
 \chi^\dagger(\vec{r})h_{\mathrm{light}}(\vec{r})\chi(\vec{r}),
 \label{eq:classD_rho_trap_theory}\\
 h_{\mathrm{light}}&\approx-\frac{i}{2}(\sigma_x\partial_x+\sigma_z\partial_y) +\left[(\alpha - \alpha_c)+c\frac{(y-y^*)^2}{L_y^2}\right]\sigma_y, \nonumber
 \end{align}
where $\chi(\vec{r})=[\chi^\dagger(\vec{r})]^T$ is a two-component
 Majorana field and $c=2 \pi^2 \alpha_c /\rho$. The trap is quadratic, unlike the linear $\rho=0$ Majorana trap. At $\alpha = \alpha_c$, the critical state is a droplet of size $L_x \ell_{\rm trap}$ with $\ell_{\rm trap}\sim L_y^{2/3}$, i.e., $\vartheta = 2/3$ ~\cite{CampostriniVicari2009PRL_TSS,CampostriniVicari2010PRA_QTSS,endmatter}.
The single-Majorana mass exhausts the symmetry-allowed relevant operators;
weak short-range interactions are irrelevant~\cite{supp}.

The $\beta$ deformation similar to \cref{eq:Thouless_parameters_space_beta} produces the phase diagram
in \cref{fig:classD_phase}(b)~\cite{supp}.  At $\rho=0$, the trap-scaling critical lines
terminate at finite $\beta$, realizing unnecessary criticality.  For
$0<|\rho|<1$, intervening $p_x\pm i p_y$ TSc lobes appear whose phase boundary exhibits multiversality: $(z,p,\vartheta)=(1,2,2/3)$ along the sides and $(1,1,1/2)$ along the caps~\cite{supp}. 
 Finally, a defect transition corresponding to the 1+1d Ising universality class expels the trapped
Kitaev chain along
$\alpha_{\rm def}(\beta)=\sin(2\pi\beta)$ that joins the TSc lobes
for $0<|\rho|<1$, or the unnecessary-critical lines at $\rho=0$, as shown in
\cref{fig:classD_phase}(b)~\cite{KitaevMajoranaChain,supp}.
}

\medskip

\noindent \emph{\textbf{Experimental realization:}} Thouless pumping and its analogues have been experimentally realized in a diverse range of platforms, including ultracold atoms \cite{CALohse,CANakajima,CALu,CAWalter}, superconducting quantum processors \cite{ScLiu}, photonic waveguide arrays \cite{PtKraus,PtVerbin,PtCerjan}, magneto/electro-mechanical systems \cite{MechGrinberg,MechXia}, and others. The high degree of control available in these platforms suggests the possibility of spatial modulation of programmable couplings, and the realization of diabolical textures and other aspects of our work.

\medskip 
\noindent \emph{\textbf{Discussion and Outlook:}} We have shown that diabolical textures produce distinct gapped regimes separated by trap-scaling criticality, intervening phases, or defect transitions, with unnecessary criticality and multiversality in noninteracting systems. Natural future directions include  textures built from families of noninvertible states and the relevance of diabolical textures to systems with slowly varying couplings, such as moiré systems.

\medskip 
 \noindent \emph{\textbf{Acknowledgments}}: We are grateful to Sashank Singam, Nick Jones, Sthitadhi Roy and Ajit Balram for helpful discussions. The authors acknowledge the use of A.I. in this work (GPT 5.4-6.0 via ChatGPT and Codex) to clarify certain conceptual points, polish the manuscript, survey literature, and develop some of the figures. All AI-assisted edits were directed, reviewed, and verified by the authors, who take full
responsibility for the scientific content and final text.
All code used to generate the figures is available at
\url{https://doi.org/10.5281/zenodo.23193045}.

\nocite{Mermin_RevModPhys.51.591,Michel_OrderedMedia_RevModPhys.52.617,ang2018classificationtopologicaldefectstextures,ang2018highercategoricalgroupsclassification}
\makeatletter
\addtocontents{toc}{\protect\tocdepth@munge}
\makeatother
\bibliography{ref}

\clearpage

\section*{End Matter}

\noindent\textbf{Trap-scaling criticality.}
\begin{figure}[!ht]
    \centering
    \includegraphics[width=\linewidth]{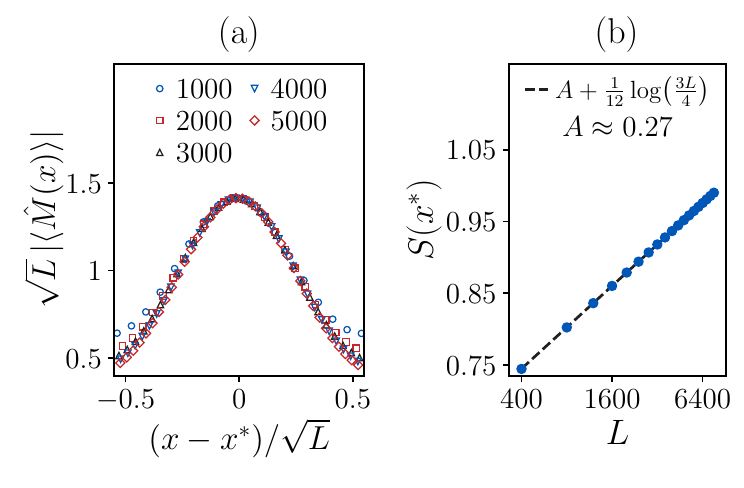}
    \caption{Finite-size scaling in the noninteracting Rice--Mele chain:
    (a) the local observable
    $\hat M(x)=\hat n_x-\hat n_{x+1}$, where
    $\hat n_x=c_x^\dagger c_x$, and (b) the von Neumann entanglement
    entropy for a cut at the trapping node $x^*=3L/4$.
    Both panels use the critical texture $\alpha=1$, $\mu=\beta=0$
    in \cref{eq:Thouless_parameters_space}. The legend in (a) gives
    the system size $L$. The scaling forms follow from the trap
    length $\ell\sim L^{1/2}$.}
    \label{fig:scaling}
\end{figure}
We briefly review the essentials of trap-scaling criticality
~\cite{CampostriniVicari2009PRL_TSS,CampostriniVicari2010PRA_QTSS,CeccarelliTorreroVicari2013PRB_CriticalParamsTSS}.
Consider a scale-invariant fixed-point theory $S_0$ perturbed by a relevant
operator $\mathcal M$ with a spatially modulated coupling,
\begin{equation}
 S=S_0+g\int d^d x\,d\tau\,
 \left(\frac{x}{L}\right)^p\mathcal M(x,\tau).
 \label{em:eq:trap_action}
\end{equation}
Here $S$ is the Euclidean action, $d$ is the spatial dimension, $\tau$ is
imaginary time, $x$ is the modulated coordinate, $L$ is the modulation
length, $g$ is its amplitude, and the positive integer $p$ specifies how
the coupling vanishes at $x=0$.
Near $x=0$, the bare coefficient of $\mathcal M$ is infinitesimal and the
system is governed by $S_0$.  Under RG it becomes of order one at a length
$\ell$ determined by
\begin{equation}
 g\frac{\ell^{p+y_{\mathcal M}}}{L^p}\sim1,\quad
 \ell\sim L^\vartheta,\quad
 \vartheta=\frac{p}{p+y_{\mathcal M}},\quad
 y_{\mathcal M}=d+z-\chi,\nonumber
\end{equation}
where $z$ is the dynamical exponent, $\chi$ is the scaling dimension of
$\mathcal M$, and $y_{\mathcal M}$ is its RG eigenvalue.  The scale $L^\vartheta$
replaces the usual infrared size: for example,
$\Delta\sim L^{-z\vartheta}$ for the characteristic trap-level spacing, and $\langle\mathcal M(x)\rangle=
 L^{-\chi\vartheta}\mathcal F(x/L^\vartheta)$ where $\mathcal F$ is a dimensionless scaling function.
For a one-dimensional conformal critical region, the Calabrese--Cardy
expression for an open-chain cut at the trapping node gives
$S=(c/6)\log\ell+\mathrm{const.}$, where $c$ is the central
charge~\cite{Cardy_Calabrese_2004,CampostriniVicari2010Entanglement}.
\Cref{fig:scaling} shows the observable and entanglement scaling for the
noninteracting Rice--Mele chain. In panel (b), the subsystem length $b=3L/4$
scales with $L$, giving $S(b)=(c\vartheta/6)\log b+\mathrm{const.}$
for this fixed cut geometry.

\medskip
\noindent\textbf{Classification.}
We briefly summarize the classification of diabolical textures. We begin
with an analogy to ordered media: here, the local
order parameter defines a continuous map $s:\mathcal M_d\to\mathcal C$ from
the sample coordinates $\mathcal M_d$ to the order-parameter space
$\mathcal C=G/H$, where $G$ and $H$
are the full and unbroken symmetry groups~\cite{Mermin_RevModPhys.51.591,Michel_OrderedMedia_RevModPhys.52.617}.  Two such maps belong to the same
homotopy class if one can be smoothly deformed into the other while remaining
in $\mathcal C$ and respecting the boundary conditions.  The classes
$[\mathcal M_d,\mathcal C]$ therefore classify topological textures~\cite{ang2018classificationtopologicaldefectstextures,ang2018highercategoricalgroupsclassification}.  For a
defect, the same construction is applied to a loop or sphere surrounding its
core: a nontrivial class cannot be extended through the enclosed region
without leaving $\mathcal C$.  This is the usual origin of vortex, hedgehog,
and skyrmion charges
~\cite{Mermin_RevModPhys.51.591,Michel_OrderedMedia_RevModPhys.52.617}.

For a diabolical texture, $\mathcal C$ is replaced by the stabilized space
$\mathscr I_d^G$ of gapped, invertible $d$-dimensional Hamiltonians with
onsite symmetry $G$, and the order parameter by the locally frozen
Hamiltonian,
$\varrho:\mathcal M_d\to\mathscr I_d^G$.  Stabilization means that adding
decoupled degrees of freedom in trivial states does not change the
classification~\cite{Gaiotto_2019,kubota2025stablehomotopytheoryinvertible}.  Although $\mathscr I_d^G$ is less pictorial than an ordinary
order-parameter space, the classes $[\mathcal M_d,\mathscr I_d^G]$ are more
tractable because the spaces $\mathscr I_d^G$ in different dimensions are
conjectured to be linked by an $\Omega$ spectrum~\cite{kitaev1,kitaev2,kitaev3}.

Kitaev's $\Omega$-spectrum relation
$\Omega\mathscr I_d^G\simeq\mathscr I_{d-1}^G$ states that closed paths in
$\mathscr I_d^G$ based at a chosen reference Hamiltonian, denoted by
$\Omega\mathscr I_d^G$, have the same
classification as $(d-1)$-dimensional invertible phases.  Repeating this
step relates a spherical $k$-parameter family to a phase in $d-k$ dimensions
~\cite{kitaev1,kitaev2,kitaev3,kubota2025stablehomotopytheoryinvertible,Xiong_2018,Gaiotto_2019}.
The full classification is
\begin{equation}
 [\varrho]\in[\mathcal M_d,\mathscr I_d^G]
 \cong h_G^d(\mathcal M_d),
 \label{em:eq:full_classification}
\end{equation}
where the brackets identify textures related by a smooth, gap-preserving
deformation, and $h_G^d$ denotes the resulting generalized-cohomology
classification~\cite{Xiong_2018,Gaiotto_2019}.  If $\mathcal M_d$ is
connected, the frozen Hamiltonian at a chosen reference point defines
the \emph{background phase}: the uniform phase obtained by taking that
Hamiltonian everywhere. The classification separates this background
from the spatial variation:
\begin{equation}
 h_G^d(\mathcal M_d)\cong\mathcal I_d^G\oplus
 \widetilde h_G^d(\mathcal M_d).
 \label{em:eq:reduced_classification}
\end{equation}
The connected components, 
$\mathcal I_d^G=\pi_0(\mathscr I_d^G)$ labels the distinct uniform
$d$-dimensional phases, while $\widetilde h_G^d$ classifies textures within
the chosen background phase~\cite{Xiong_2018,supp}.
For a spherical $k$-dimensional texture we get
\begin{equation}
 \begin{gathered}
 \widetilde h_G^d(S^k)\cong\mathcal I_{d-k}^G,
 \qquad 1\leq k\leq d,\\[-2pt]
 \varrho_k:\pi_k(\mathcal M_d)\longrightarrow\mathcal I_{d-k}^G.
 \end{gathered}
 \label{em:eq:spherical_classification}
\end{equation}
Here $S^k$ is a $k$-sphere, $\pi_k(\mathcal M_d)$ labels the distinct
spherical cycles in the sample, and $\varrho_k$ assigns to each one the
trapped $(d-k)$-dimensional phase.  This gives the charge of the Rice--Mele
loop and the Kitaev phase of the class D loop.  Spheres do not probe every
sample geometry or how lower-dimensional phases fit together where regions
meet~\cite{Gaiotto_2019,supp}.  For example, on the spatial torus
$T^2$, $h_G^2(T^2)$ contains an $\mathcal I_0^G$ contribution even though
$\pi_{k\geq2}(T^2)=0$, so it is invisible to every $\varrho_{k\geq2}$.
The Supplemental Material discusses the classification and this example in more
detail~\cite{supp}.

\medskip
\noindent\textbf{Constructing examples.}
We next describe two microscopic routes to diabolical textures.  

\textit{Suspension.} We build a higher-dimensional family by stacking a
lower-dimensional family and its inverse along a new spatial direction,
then coupling neighboring layers into trivial pairs in two inequivalent
dimerization patterns
~\cite{kitaev1,kitaev2,kitaev3,Wenetal_topologicalfamilies,AP_ThoulessPhasediagrams}.
The Rice--Mele pump itself can be interpreted as the suspension of a zero-dimensional charge;
the two pairing patterns become the two bond dimerizations of the chain.
Repeating the construction gives families over $S^d$ in $d$ spatial
dimensions, whose spatial embedding produces $d$-dimensional diabolical
textures~\cite{AP_ThoulessPhasediagrams,supp}.

For a concrete two-dimensional example, we suspend the Rice--Mele pump by
stacking chains with alternating pump orientation along a second direction.  The two
dimerizations of the interchain coupling pair each chain with an inverse and
contract the stack in two inequivalent ways.  Interpolating between them adds
the parameter $\delta_2$ to the Rice--Mele parameters $(\delta_1,J)$.  The
resulting two-dimensional Hamiltonian is~\cite{AP_ThoulessPhasediagrams}.
\begin{multline}
 H_{\rm RM}^{[2]}=\sum_{\vec{x}}\sum_{a=1}^2
 \left(t^a_{\vec{x}}c^\dagger_{\vec{x}}c_{\vec{x}+\hat{\vec{e}}_a}+\mathrm{h.c.}\right)
 +\sum_{\vec{x}}J_{\vec{x}}c^\dagger_{\vec{x}}c_{\vec{x}},\\
 t^1_{\vec{x}}=(-1)^{x_2}\frac{1-(-1)^{x_1}\delta_1}{2},\qquad
 t^2_{\vec{x}}=\frac{1+(-1)^{x_2}\delta_2}{2},\\
 J_{\vec{x}}=(-1)^{x_1+x_2}J .
 \label{em:eq:RM_2d_ascendant}
\end{multline}
Here $\vec{x}=(x_1,x_2)$ labels square-lattice sites and $\hat{\vec{e}}_a$ is a
unit lattice vector.  At fixed
$\delta_1^2+\delta_2^2+J^2=R^2$, this defines a gapped family over $S^2$
surrounding the diabolical point $\delta_1=\delta_2=J=0$.  For a periodic sample, this can be converted to a diabolical texture of degree $+1$ using the
Wilson--Dirac map~\cite{QiWuZhang2006}
\begin{equation}
 \begin{pmatrix}
  \delta_1(\vec{r})\\ \delta_2(\vec{r})\\ J(\vec{r})
 \end{pmatrix}
 =\frac{R}{\mathcal N(\vec{u})}
 \begin{pmatrix}
  -\sin u_1\\ \sin u_2\\ m_{\rm W}-\cos u_1-\cos u_2
 \end{pmatrix},
 \label{em:eq:RM_2d_texture}
\end{equation}
where $\vec{r}=(r_1,r_2)$ is the slowly varying spatial coordinate,
$u_a=2\pi r_a/L_a$, $L_a$ is the system length in direction $a$,
$0<m_{\rm W}<2$ is the Wilson mass, and $\mathcal N$ is the norm of the
displayed vector.  This map covers the parameter sphere once and
therefore converts the family into a diabolical, skyrmion-like texture
~\cite{Ryu,AP_ThoulessPhasediagrams}.

The class D Hamiltonian in the main text is
obtained by applying the first suspension to the Kitaev chain: the two
inequivalent interlayer dimerizations provide the two contractions joined by
the loop~\cite{KitaevMajoranaChain}.  Further details of the suspension
procedure are given in the Supplemental Material~\cite{supp}.

\vspace{4pt}
\noindent\textit{Pivoting.}
A complementary microscopic route is pivoting
~\cite{NatVerresenThorngren_Pivots_10.21468/SciPostPhys.14.2.012,Shiozaki_SPTPUmp_PhysRevB.106.125108,APNickPaul_ChiralClock,JonesThorngrenAP_SPTPump}.
It can generate gapped loops that trap an invertible phase in one lower dimension. For a reference $H_0$ and a local pivot
$H_{\rm pivot}=\sum_{x=1}^Lh_{\rm p}(x)$, it produces the isospectral family
\begin{equation}
 H(\theta)=e^{i\theta H_{\rm pivot}}H_0e^{-i\theta H_{\rm pivot}}.
 \label{em:eq:pivot_loop}
\end{equation}
Here $h_{\rm p}(x)$ is a local Hamiltonian term and $\theta$ is the family
parameter.  Writing the resulting family as $H(\theta)=\sum_jh_j(\theta)$,
we introduce the spatial angle in each local term to obtain a system with diabolical texture
\begin{equation}
 H_{\rm tex}=\sum_{j=1}^Lh_j(\theta_j),\qquad
 \theta_j=\frac{2\pi j}{L}.
 \label{em:eq:pivot_texture}
\end{equation}
Here $j$ labels sites and $L$ is the system length.  For example,
$H_0=-\sum_{j=1}^L\sigma_{x,j}$ and
$H_{\rm pivot}=-(1/4)\sum_{j=1}^L\sigma_{z,j}\sigma_{z,j+1}$ generate the Ising
charge-pump loop $H_{\rm Ising}(\theta)$, protected by the global $\mathbb Z_2$ spin-flip symmetry
$P=\prod_j\sigma_{x,j}$, where $\sigma_{x,j},\sigma_{y,j},\sigma_{z,j}$ are Pauli operators on site
$j$, with periodic spin indices.  The texture Hamiltonian is
\begin{multline}
 H_{\rm tex}=\sum_{j=1}^L\bigl[
 \sin^2\left( \frac{\pi j}{L} \right)\,\sigma_{z,j-1}\sigma_{x,j}\sigma_{z,j+1}
 -\cos^2\left( \frac{\pi j}{L} \right)\,\sigma_{x,j}\\
 -\sin\left( \frac{\pi j}{L} \right) \cos\left( \frac{\pi j}{L} \right)(\sigma_{z,j-1}\sigma_{y,j}
 +\sigma_{y,j}\sigma_{z,j+1})\bigr].
 \label{em:eq:Ising_pivot_texture}
\end{multline}
For a slowly varying unit winding, its ground state carries odd
$\mathbb Z_2$ charge~\cite{Shiozaki_SPTPUmp_PhysRevB.106.125108}.
Onsager-integrable and group-cohomology pivots provide further examples~\cite{JonesThorngrenAP_SPTPump}.

The two constructions can also be composed.  For example, suspending the
Ising pump gives a gapped family over a two-sphere.  A degree-one spatial
embedding in a two-dimensional system with Ising symmetry produces a
$\mathbb Z_2$-classified, skyrmion-like diabolical texture that traps the
nontrivial Ising charge
~\cite{Wenetal_topologicalfamilies,JonesThorngrenAP_SPTPump}.

\clearpage
\onecolumngrid
\raggedbottom
\counterwithout*{equation}{section}
\setcounter{section}{0}
\setcounter{secnumdepth}{3}
\setcounter{subsection}{0}
\setcounter{subsubsection}{0}
\setcounter{equation}{0}
\setcounter{figure}{0}
\setcounter{table}{0}
\renewcommand{\thesection}{\Roman{section}}
\renewcommand{\thesubsection}{\Alph{subsection}}
\renewcommand{\thesubsubsection}{\arabic{subsubsection}}
\renewcommand{\theequation}{S\arabic{equation}}
\renewcommand{\thefigure}{S\arabic{figure}}
\renewcommand{\thetable}{S\Roman{table}}
\renewcommand{\theHsection}{supp.\arabic{section}}
\renewcommand{\theHsubsection}{supp.\arabic{section}.\arabic{subsection}}
\renewcommand{\theHsubsubsection}{supp.\arabic{section}.\arabic{subsection}.\arabic{subsubsection}}
\renewcommand{\theHequation}{supp.\arabic{equation}}
\renewcommand{\theHfigure}{supp.\arabic{figure}}
\renewcommand{\theHtable}{supp.\arabic{table}}
\begin{center}
\Large\textbf{Supplemental Material}
\end{center}
\setcounter{tocdepth}{2}
\makeatletter
\addtocontents{toc}{\protect\tocdepth@restore}
\makeatother
\tableofcontents

\section{Rice--Mele texture: additional details}

\label{app:Landau Levels}
\subsection{Dirac trap and Landau-level correspondence}
This section contains additional details of the Rice--Mele texture studied in
the main text.  Let us write down the Hamiltonian of the Rice--Mele model~\cite{RMmodel} 
\begin{equation}
    \Ham_{\text{RM}} = \sum_{x=1}^L\left[ \frac{\left(1-(-1)^x \delta(x) \right)}{2}(c^\dagger_x c_{x+1} + h.c.) + (-1)^x J(x) c^\dagger_x c_x\right]\label{smeq:H_Rice_Mele}
\end{equation}
Here $c_x$ annihilates a spinless fermion on site $x$, and $L$ is
the number of sites. We set the nearest-neighbor site spacing and overall
hopping scale to one. Each two-site unit cell has length two. The textured Hamiltonian is obtained by varying the parameters as
\begin{equation}
    \delta(x) =\cos\left(\frac{2 \pi x}{L} (1-\beta) \right), ~J(x) =\alpha + \sin\left(\frac{2 \pi x}{L} (1-\beta) \right) \label{smeq:Thouless_parameters_space}
\end{equation}
 At each $x$, let
$|u_-(k;x)\rangle$ be the occupied state of the homogeneous Bloch
Hamiltonian obtained by freezing $\delta(x)$ and $J(x)$. Its Berry phase and winding are 
\begin{equation}
 \gamma(x)=\oint_{\rm BZ}dk\,
 i\langle u_-(k;x)|\partial_k u_-(k;x)\rangle\pmod{2\pi},
 \label{smeq:RM_local_Berry_phase}
\qquad
\nu_\gamma= \frac{1}{2\pi}\int_0^Ldx\,\partial_x\gamma(x).
\end{equation}
We work in the reduced Brillouin zone corresponding to $-\pi/2<k\leq\pi/2$.

\subsubsection{Dirac Landau-level correspondence}
\begin{figure}[H]
    \centering
    \includegraphics[width=\linewidth]{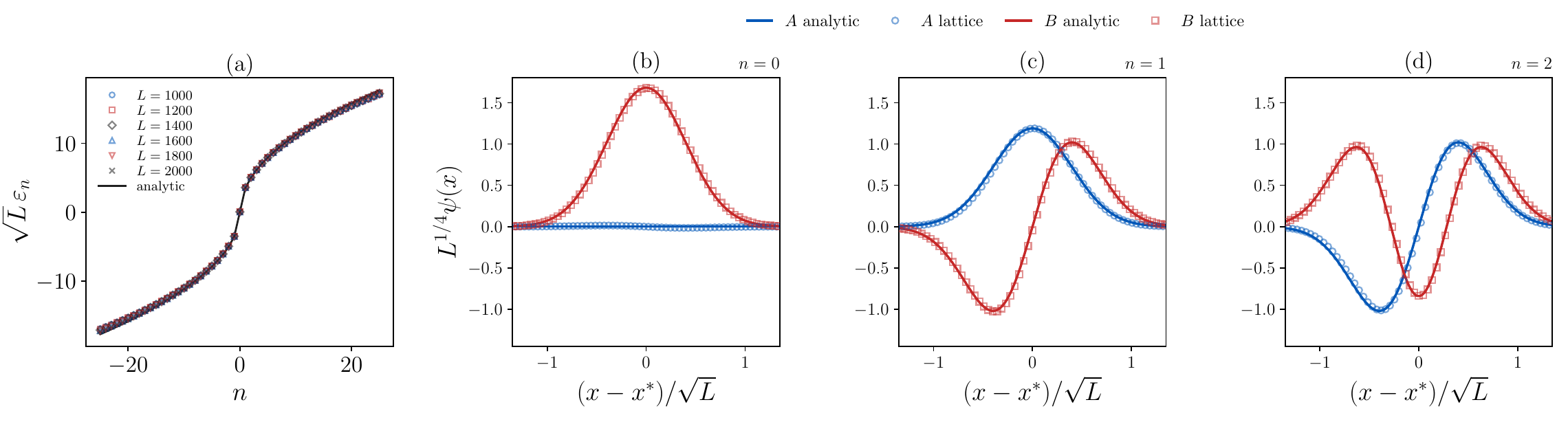}
    \caption{Lattice and continuum comparison at the Rice--Mele
    trap-critical point $\alpha=1$, $\beta=\mu=0$, with $x^*=3L/4$.
    We diagonalize \cref{smeq:H_Rice_Mele}.
    (a) The central level and 25 levels on either side 
    compared with $\sqrt{L}\,\varepsilon_n=
    \operatorname{sgn}(n)\sqrt{4\pi|n|}$ from
    \cref{smeq:Landau_energies}. The level closest to zero is labeled
    $n=0$: increasing and decreasing energies give positive and negative
    indices. (b--d) Signed sublattice amplitudes of the $n=0,1,2$ modes
    at $L=2000$. Open circles and squares show the odd-site ($A$) and
    even-site ($B$) lattice components after removing $(-1)^j$ and
    multiplying by $L^{1/4}$, as defined in \cref{smeq:remove_sublattice}. }
    \label{fig:RM_Landau_comparison}
\end{figure}
The low-energy description of \cref{smeq:H_Rice_Mele} near the
trap-scaling critical points
$(\alpha^*,x^*)=(1,3L/[4(1-\beta)])$ and
$(-1,L/[4(1-\beta)])$ has the form
\begin{equation}
 \Ham_{\text{RM}} \approx \int dx~ \psi^\dagger(x) \left( i \sigma_2   \partial_x 
 +\text{sgn}(\alpha^*) \frac{(x -x^*)}{\ell^2} \sigma_1  - (\alpha - \alpha^*)  \sigma_3 \right) \psi(x), \quad \ell = \sqrt{\frac{L}{2 \pi(1-\beta)}} 
 \label{smeq:Dirac_trapnode}
\end{equation} 
Here $\psi=(\psi_A,\psi_B)^{\mathsf T}$ is the continuum
two-sublattice fermion field and $\sigma_{1,2,3}$ are Pauli matrices in
sublattice space. The length $\ell$ is the width of the trapped mode and,
for fixed $\beta<1$, scales as $L^{1/2}$.
The first quantized Hamiltonian of \cref{smeq:Dirac_trapnode} is equivalent to the Landau-level problem of a 2d massive Dirac fermion in magnetic field. We present the details of this equivalence here. Consider a 2d Dirac fermion with constant mass $m$ and velocity $v$ in a uniform magnetic field $\vec{B} = B \hat{\vec{e}}_z$. This has a Hamiltonian~\cite{SharapovGusyninBeck2004,Goerbig2011}
\begin{equation}
    h = v\left( \Pi_x \sigma_2 + \Pi_y \sigma_1  \right) - m \sigma_3
\end{equation}
Here, $\vec{\Pi} = -i \vec{\nabla} + \vec{A}$, where $\vec{A}$ is a vector potential satisfying $\vec{B} = \vec{\nabla}\times\vec{A}$. We start by considering the Landau gauge $\vec{A} = Bx \hat{\vec{e}}_y$, where the Hamiltonian becomes
\begin{equation}
h =  (-i v\partial_x \sigma_2 + v(-i \partial_y+Bx) \sigma_1) - m \sigma_3
\end{equation}
In this gauge, $h$ has translation symmetry only in y-direction, and therefore its eigenstates are of the form 
\begin{equation}
    \Psi_{n,k}(x,y) = \frac{e^{iky}}{\sqrt{L_y}}\psi_{n,k}(x)
\end{equation}
where $L_y$ is the auxiliary length in the
translation-invariant direction and $\psi_{n,k}$ are eigenstates of
\begin{equation}
\label{eq:Landau Gauge Hamiltonian}
h_k = -i  v\partial_x \sigma_2 + v\operatorname{sgn}(B)\frac{(x-x^*)}{\ell^2}\sigma_1 - m \sigma_3,~\quad \ell = \sqrt{\frac{1}{|B|}},~x^* = -\text{sgn}(B)~k \ell^2
\end{equation}
$\ell$ is the magnetic length and $x^*$ the guiding center.
Under the correspondence with \cref{smeq:Dirac_trapnode},
the velocity is $v=1$, the magnetic length is
$\ell=\sqrt{L/[2\pi(1-\beta)]}$, the Dirac mass is
$m=\alpha-\alpha^*$, $\operatorname{sgn}(B)=-\operatorname{sgn}(\alpha^*)$,
and $k=-\operatorname{sgn}(B)x^*/\ell^2$ fixes the guiding center at the
trap node.
Conjugating $h_k$ by $\sigma_3$ produces the first-quantized
Hamiltonian in \cref{smeq:Dirac_trapnode}. The
eigenvalues of $h_k$ are~\cite{SharapovGusyninBeck2004}
\begin{equation}
   \varepsilon_0 =  -\operatorname{sgn}(B)m,\;\; \varepsilon_{\pm n} = \pm  \sqrt{\frac{2v^2}{\ell^2}n+m^2}\;\;\text{for } n \in \mathbb{N} \label{smeq:Landau_energies}
\end{equation}
We use the trap-scaling relation reviewed in the End Matter,
\begin{equation}
 \vartheta=\frac{p}{p+y_M},\qquad
 \xi\sim L^\vartheta,\qquad
 \Delta_{\rm trap}\sim L^{-z\vartheta}.
 \label{smeq:trap_scaling}
\end{equation}
Here $p$ is the order of the spatial zero of the modulated coupling,
$y_M$ is its renormalization-group (RG) eigenvalue, $z$ is the dynamical
exponent, and $\xi$ is the size of the critical region. For the free
Dirac and Majorana masses, $y_M=z=1$. For the free quadratic band-edge
detuning, $y_M=z=2$. Thus these free theories have
$\vartheta=p/(p+z)$
~\cite{CampostriniVicari2009PRL_TSS,CampostriniVicari2010PRA_QTSS}.

Thus the Rice--Mele trap has
$\varepsilon_0=\operatorname{sgn}(\alpha^*)m$ and
$\varepsilon_{\pm n}=\pm\sqrt{m^2+2n/\ell^2}$. At $m=0$, its transverse
width and level spacing scale as $\ell\sim L^{1/2}$ and
$\Delta\sim\ell^{-1}\sim L^{-1/2}$. Equivalently, the Dirac mass has
$z=1$ and a linear spatial zero, $p=1$, so
\cref{smeq:trap_scaling} gives $\vartheta=1/2$.
In the Rice--Mele discussion, ``gap'' refers to the neutral excitation
gap at fixed particle number $N$. For ordered single-particle energies
$\varepsilon_1\leq\varepsilon_2\leq\cdots$, it is
$\Delta_N=\varepsilon_{N+1}-\varepsilon_N$ and scales as
$\Delta_N\sim L^{-1/2}$ at the critical point.
For $B>0$, the eigenfunctions $\psi_n$ are given by:
\begin{equation}
\label{smeq: Landau Level wfns B > 0}
\psi_0 (x) = \begin{pmatrix}
  \varphi_0 (x) \\ 0
\end{pmatrix} ,\quad \psi_{\pm n}(x) =  \begin{pmatrix}
     \xi_{ n}^\mp \;\varphi_{n} (x)\\ \pm \xi_{ n}^\pm \;\varphi_{n-1} (x)
\end{pmatrix} \;\; \text{for } n \in \mathbb{N},
\quad \xi_{ n}^\pm =\sqrt{\frac{|\varepsilon_n| \pm m}{2|\varepsilon_n|}} 
\end{equation}
where $\varphi_n$ are harmonic oscillator eigenfunctions
\begin{equation}
\varphi_n (x) = \frac{1}{\sqrt{2^n n!}} \frac{1}{\sqrt{\sqrt{\pi}\ell}} ~ H_n \left(\frac{x-x^*}{\ell} \right) ~ \exp\left(-\frac{(x-x^*)^2}{2\ell^2}\right)
\end{equation}
and $H_n(x)$ are Hermite polynomials. Similarly, for the case $B < 0$, the eigenfunctions are:
\begin{equation}
\label{smeq: Landau Level wfns B < 0}
\psi_0 (x) = \begin{pmatrix}
 0\\ \varphi_0 (x) 
\end{pmatrix} ,\quad \psi_{\pm n}(x) =  \begin{pmatrix}
    \mp \xi_{n}^\mp \;\varphi_{n-1} (x)\\ \xi_{n}^\pm \;\varphi_n (x)
\end{pmatrix} \;\; \text{for } n \in \mathbb{N}
\end{equation}
We compare these continuum states with the lattice eigenstates
at $\alpha=1$, $\beta=0$ in \cref{fig:RM_Landau_comparison}. Let
$\psi_n^{\mathrm{lat}}(x)$ be a real, normalized eigenvector of
\cref{smeq:H_Rice_Mele}, with $\sum_x|\psi_n^{\mathrm{lat}}(x)|^2=1$.
The odd and even sites form sublattices $A$ and $B$, respectively.
At $k=\pi/2$, the low-energy wavefunction contains the Bloch factor
$e^{2ikj}=(-1)^j$ in the two-site cell $j$. Removing this factor exposes the slowly varying
amplitudes on the two sublattices. In panels (b--d) we plot
\begin{equation}
 L^{1/4}(-1)^j
 \begin{pmatrix}
 \psi_n^{\mathrm{lat}}(2j+1)\\
 \psi_n^{\mathrm{lat}}(2j+2)
 \end{pmatrix},\qquad j=0,\ldots,L/2-1. \label{smeq:remove_sublattice}
\end{equation}
The Gaussian zero mode and the successive oscillator profiles agree
with the lattice data, including their relative sublattice signs.

\subsection{Finite chemical potential: metallic droplets and trap scaling}
\label{app:RM_finite_mu}

We now show how a finite chemical potential broadens 
isolated trap-critical lines into metallic lobes, determine the resulting
trap-scaling regimes, and derive various aspects of the finite-$\beta$ phase diagram.

\subsubsection{Frozen spectrum and metallic droplets}

For the Hamiltonian and spatial profile in
\cref{smeq:H_Rice_Mele,smeq:Thouless_parameters_space},
a uniform chemical potential gives
\begin{equation}
 \Ham_\mu=\Ham_{\rm RM}-\mu\sum_x c_x^\dagger c_x .
 \label{smeq:RM_mu_Hamiltonian}
\end{equation}
We take $0<\mu<1$ below corresponding to electron doping. The hole-doped bulk results can be obtained by
replacing $\mu$ by $|\mu|$.  Unlike the two Dirac masses $\delta$ and $J$,
$\mu$ multiplies the identity and shifts the single-particle spectrum
without changing its eigenfunctions. To obtain the local band edge directly from the lattice model, freeze
$\delta(x)$ and $J(x)$ at a fixed position and group the odd and even sites
as $A_j=c_{2j-1}$ and $B_j=c_{2j}$. Using Fourier factors $e^{2ikj}$ for
both sublattices and writing $t_1=(1+\delta)/2$, $t_2=(1-\delta)/2$,
the two-site-cell Bloch Hamiltonian is
\begin{equation}
 \begin{aligned}
 h_{\mathrm{RM}}(k;x)&=
 \begin{pmatrix}
  -J(x)&t_1(x)+t_2(x)e^{-2ik}\\
  t_1(x)+t_2(x)e^{2ik}&J(x)
 \end{pmatrix},\\
 \varepsilon_\pm(k,x)&=\pm\sqrt{J^2(x)+\cos^2 k
 +\delta^2(x)\sin^2 k} .
 \end{aligned}
 \label{smeq:RM_mu_exact_cell_bloch}
\end{equation}
The translation between unit cells has length two, giving the phases
$e^{\pm2ik}$. The band minimum is at $k=\pi/2$. Below,
$q=k-\pi/2$ denotes momentum measured from it. The
corresponding energies of $\Ham_\mu$ are $\varepsilon_\pm(k,x)-\mu$.
For the texture in \cref{smeq:Thouless_parameters_space}, we write the
spatially varying angle as $\zeta_\beta(x)=2\pi(1-\beta)x/L$.
Since $|\delta(x)|\leq1$, the local conduction-band bottom and its
small-$q$ expansion are
\begin{equation}
 \varepsilon_c(x)=\min_k\varepsilon_+(k,x)
 =R(x)=\sqrt{J^2(x)+\delta^2(x)},\qquad
 \varepsilon_+^2(\pi/2+q,x)=R^2(x)+v^2(x)q^2
 +\mathcal O(q^4),\quad
 v^2(x)=1-\delta^2(x).
 \label{smeq:RM_mu_exact_band_edge}
\end{equation}
A point is therefore locally metallic precisely when $R(x)<\mu$.

\subsubsection{Droplet geometry at $\beta=0$}

\begin{figure}[tbp]
 \centering
 \includegraphics[width=\linewidth]{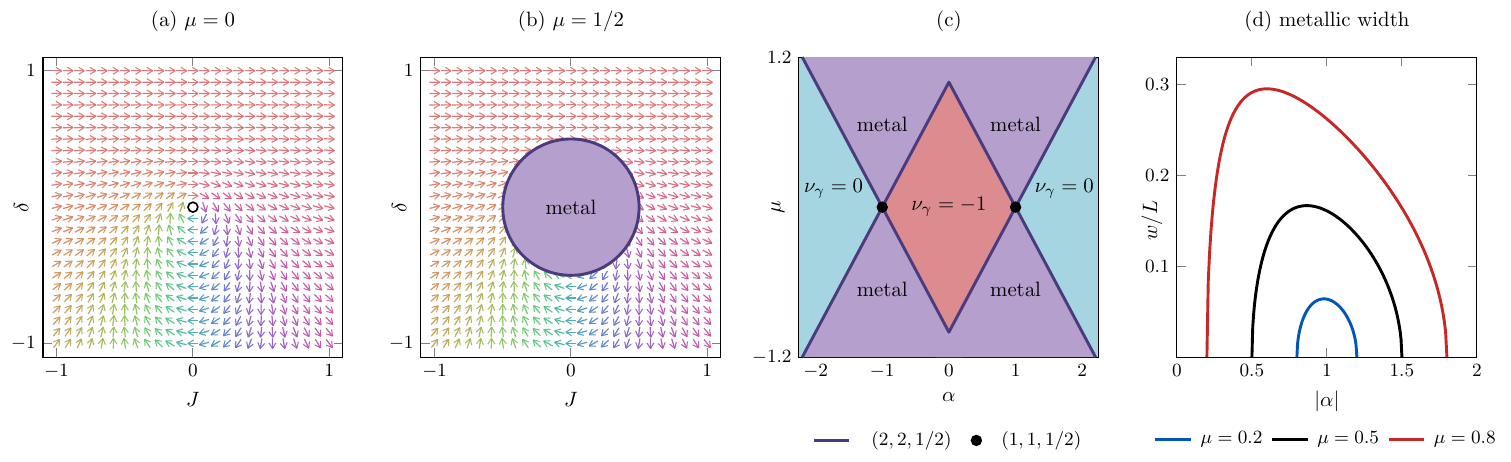}
 \caption{Rice--Mele phase diagram and metallic-droplet
 width. (a) At $\mu=0$, the system is a gapped insulator away from the
 band-touching point $J=\delta=0$, around which the Berry-phase texture
 winds. (b) At $\mu=1/2$, the region
 $R=\sqrt{J^2+\delta^2}<\mu$ is metallic and is separated from the
 surrounding insulator by the critical circle $R=\mu$.
 (c) Phase diagram of the $\beta=0$ spatial texture in the
 $(\alpha,\mu)$ plane for $|\mu|<1$. Beyond this range, the shading
 shows the continuum extrapolation. The straight phase boundaries
 $|\alpha|=1\pm|\mu|$ follow from \cref{smeq:RM_mu_lobes}. Inside the
 shaded metallic regions, an extensive part of the chain is metallic.
 The remaining regions are gapped and labeled by $\nu_\gamma$.
 Legend entries give $(z,p,\vartheta)$: the lines have the quadratic
 band-edge trap in \cref{smeq:RM_mu_quadratic_scaling}, while the black
 points $(\alpha,\mu)=(\pm1,0)$ have the linear Dirac trap of
 \cref{smeq:Dirac_trapnode}. Negative $\mu$ gives hole-doped droplets.
 (d) For the same $\beta=0$ spatial texture, the metallic fraction $w/L$ from
 \cref{smeq:RM_mu_exact_width}. It is nonzero for
 $1-\mu<|\alpha|<1+\mu$.}
 \label{fig:RM_homogeneous_droplet}
\end{figure}

At $\beta=0$, the frozen positive band edge is
\begin{equation}
 R^2(x)=\delta^2(x)+J^2(x)
 =1+\alpha^2+2\alpha\sin\frac{2\pi x}{L},
 \qquad R_{\min}=\bigl||\alpha|-1\bigr| .
 \label{smeq:RM_mu_R}
\end{equation}
A point is locally metallic when $R(x)<\mu$.  Thus a metallic droplet
exists for
\begin{equation}
 1-\mu<|\alpha|<1+\mu .
 \label{smeq:RM_mu_lobes}
\end{equation}
\Cref{fig:RM_homogeneous_droplet}(c) shows this phase diagram in the
$(\alpha,\mu)$ plane, including hole doping. The gapped central sector
$|\alpha|<1-|\mu|$ has $\nu_\gamma=-1$, while
$|\alpha|>1+|\mu|$ has $\nu_\gamma=0$: both follow by reducing $\mu$
to zero without closing the local gap. Define
\begin{equation}
 \zeta^*=\begin{cases}
  3\pi/2,&\alpha>0,\\
  \pi/2,&\alpha<0,
 \end{cases}
 \qquad \phi_\mu=\arccos\!\left(
 \frac{1+\alpha^2-\mu^2}{2|\alpha|}\right).
 \label{smeq:RM_mu_angle}
\end{equation}
The metallic interval is
$\zeta^* -\phi_\mu<2\pi x/L<\zeta^*+\phi_\mu$, with width
\begin{equation}
 w=\frac{L}{\pi}\phi_\mu .
 \label{smeq:RM_mu_exact_width}
\end{equation}
In particular, on $|\alpha|=1$ the exact width is
\begin{equation}
 w=\frac{2L}{\pi}\arcsin\frac{\mu}{2}.
 \label{smeq:RM_mu_center_width}
\end{equation}
At fixed parameters inside a metallic lobe,
$w/L$ approaches a nonzero constant.  The droplet contains
$\mathcal O(L)$ low-energy states and has a global level spacing of order
$L^{-1}$. \Cref{fig:RM_homogeneous_droplet}(a,b) shows the
homogeneous phase diagrams, and panel (d) plots the resulting extensive
droplet width.

\subsubsection{Trap-scaling regimes}

The metal--insulator phase boundary has a different critical theory from
the original relativistic trap. The exact frozen-band expansion in
\cref{smeq:RM_mu_exact_band_edge} gives, near $R(x)=\mu$,
\begin{equation}
 \varepsilon_+(\pi/2+q,x)-\mu=\frac{v^2(x)q^2}{2\mu}+R(x)-\mu
 +\mathcal O(q^4).
 \label{smeq:RM_mu_local_band_expansion}
\end{equation}
The trap order $p$ is therefore the order of the first nonzero spatial
derivative of $R$ at the point where the band reaches the Fermi level. First consider the threshold at which $\varepsilon_c(x)$ reaches $\mu$ at
a spatial minimum $x^*$ of the local band edge $R(x)$ inside the chain. This will describe the
vertical portions of the small-$\beta$ phase boundary derived below. Let
$X=x-x^*$ and let $\alpha_c$ be the corresponding value of $\alpha$. At
the threshold, $R(x^*;\alpha_c)=\mu$ and
$\partial_xR(x^*;\alpha_c)=0$. For nearby $\alpha$, define the detuning
$t=\mu-R(x^*;\alpha)=\mu-\bigl||\alpha|-1\bigr|$: hence $t=0$ at
$\alpha=\alpha_c$, $t>0$ inside the metallic lobe, and $t<0$ outside.
Expanding about this minimum and replacing $q\to-i\partial_X$, we obtain
\begin{equation}
 \begin{aligned}
 R(x)-\mu&=-t+\frac{|\alpha_c|}{2\mu}
 \left[\frac{2\pi(1-\beta)X}{L}\right]^2+\cdots,\\
 h_{\mathrm{minimum}}&=-\frac{\partial_X^2}{2\mu}-t+\frac{|\alpha_c|}{2\mu}
 \left[\frac{2\pi(1-\beta)X}{L}\right]^2+\cdots .
 \end{aligned}
 \label{smeq:RM_mu_quadratic_trap}
\end{equation}
Here $\delta(x^*)=0$, and hence $v(x^*)=1$.   The factor
$|\alpha_c|$ is a nonuniversal curvature: it does not affect
the powers of $L$. The leading spatial variation is quadratic because
$R'(x^*)=0$ at the minimum.
This is a $z=2$ band edge in a quadratic trap: $p=2$ and
\begin{equation}
 \vartheta=\frac{p}{p+2}=\frac12,\qquad
 \xi_{\rm min}=\sqrt{\frac{L}{2\pi(1-\beta)}}\,|\alpha_c|^{-1/4}
 \propto L^{1/2},\qquad
 \Delta_{\rm min}\propto L^{-z\vartheta}=L^{-1} .
 \label{smeq:RM_mu_quadratic_scaling}
\end{equation}
Thus the trap-scaling exponent happens to remain $1/2$, but the dynamical
exponent and the energy scaling differ from the $\mu=0$ Dirac trap,
as confirmed by the exact-diagonalization results in
\cref{fig:RM_unwinding}(c).

\begin{figure}[H]
 \centering
 \includegraphics[width=.98\linewidth]{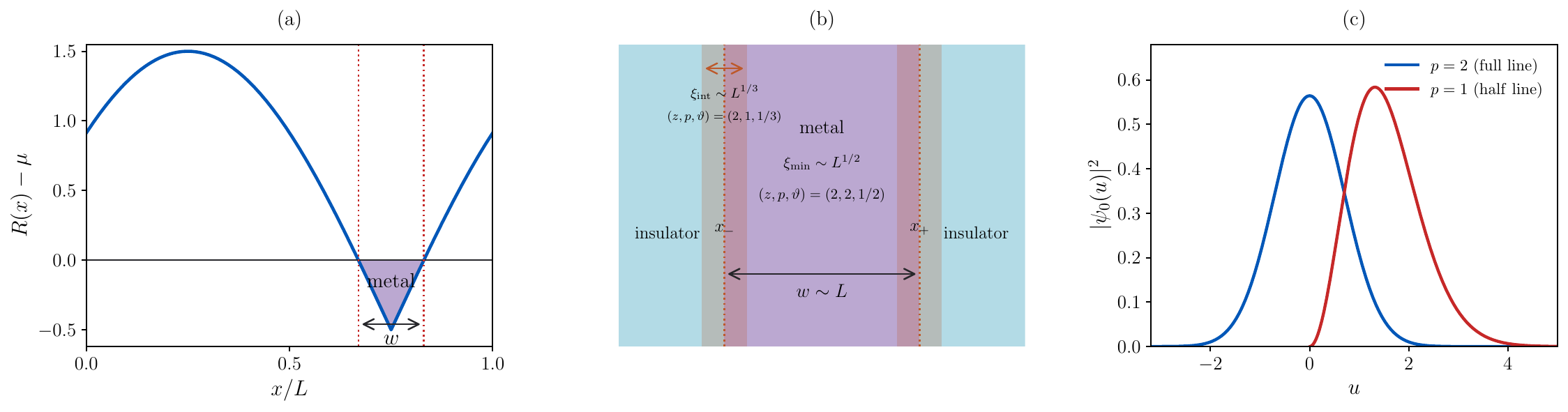}
 \caption{Metallic droplet and its trap scales for
 $\alpha=1$, $\mu=1/2$, and $\beta=0$.
 (a) The curve $R(x)-\mu$ is evaluated analytically from
 \cref{smeq:RM_mu_R}: its zeros and the shaded interval $R(x)<\mu$ give
 $x_\pm$ and the extensive width $w=x_+-x_-$ in
 \cref{smeq:RM_mu_exact_width}.
 (b) Schematic separation of scales inside a developed droplet. The
 droplet has $w=\mathcal O(L)$, while each linear turning region has
 $(z,p,\vartheta)=(2,1,1/3)$ and width
 $\xi_{\rm int}\sim L^{1/3}$, with
 $\Delta_{\rm int}\sim L^{-2/3}$ [\cref{smeq:RM_mu_linear_scaling}]. At a
 lobe onset the two interfaces merge, $w\to0$, and the spatial
 minimum of $R(x)$ gives $(z,p,\vartheta)=(2,2,1/2)$ and
 $\xi_{\rm min}\sim L^{1/2}$ and $\Delta_{\rm min}\sim L^{-1}$
 [\cref{smeq:RM_mu_quadratic_scaling}].
 (c) Normalized lowest-mode densities in scaled coordinates:
 $u=X/\xi_{\rm min}$ for the $p=2$ full-line trap, where $X=x-x^*$,
 and $u=s/\xi_{\rm end}$ for the $p=1$ half-line trap, where $s=L-x$.
 The corresponding lengths are given in
 \cref{smeq:RM_mu_quadratic_scaling,smeq:RM_mu_cap_theory}.
 The $p=2$ curve is the analytic harmonic-oscillator density
 $e^{-u^2}/\sqrt\pi$ obtained by rescaling
 \cref{smeq:RM_mu_quadratic_trap}. The $p=1$ curve is obtained by
 finite-difference diagonalization of the dimensionless half-line
 operator $-\partial_u^2+u$ on $0<u<10$ with 900 intervals and a
 Dirichlet condition at $u=0$. It is the rescaled endpoint theory in
 \cref{smeq:RM_mu_cap_theory}.}
 \label{fig:RM_droplet_traps}
\end{figure}

Inside a developed droplet, let $x_c$ be either point at which
$R(x_c)=\mu$. The
conduction-band edge varies linearly with the spatial displacement $x-x_c$ from the interface,
\begin{equation}
 h_{\mathrm{interface}}=-\frac{v_c^2\partial_x^2}{2\mu}
 +g_c(x-x_c)+\cdots,\qquad
 v_c^2=1-\delta^2(x_c),\qquad \zeta_c=\zeta_\beta(x_c),\qquad
 |g_c|=\left|\partial_xR(x_c)\right|
 =\frac{2\pi(1-\beta)}{L}\frac{|\alpha\cos\zeta_c|}{\mu} .
 \label{smeq:RM_mu_linear_edge}
\end{equation}
This is a $z=2$ linear turning point, so $p=1$ and
\begin{equation}
 \vartheta_{\rm int}=\frac13,\qquad
 \xi_{\rm int}=\left(\frac{v_c^2}{2\mu|g_c|}\right)^{1/3}
 \propto L^{1/3},\qquad
 \Delta_{\rm int}\propto L^{-2/3} .
 \label{smeq:RM_mu_linear_scaling}
\end{equation}
The interface crossover is narrower than the extensive metallic region. Its
$L^{-2/3}$ scale does not replace the smaller $L^{-1}$ level spacing of
the metallic interior.  These results follow from the general
trap-scaling relation in \cref{smeq:trap_scaling}, with $z=2$ and
$p=1$ at an interface
~\cite{CampostriniVicari2009PRL_TSS,CampostriniVicari2010PRA_QTSS,CeccarelliTorreroVicari2013PRB_CriticalParamsTSS}. The real-space droplet, its interface crossover scales, and
representative lowest modes of the $p=2$ and half-line $p=1$ trap
Hamiltonians are shown in
\cref{fig:RM_droplet_traps}. The half-line trap describes a droplet
disappearing at the open end, as derived below in
\cref{smeq:RM_mu_cap_theory}.

\subsubsection{Finite-$\beta$ phase diagram and lobe phase boundaries}

As $\beta$ varies, the spatial profile in \cref{smeq:Thouless_parameters_space} covers
$0\leq\zeta_\beta(x)\leq2\pi(1-\beta)$. As long as the
minimum of $R(x)$ remains inside the chain, the threshold
$R_{\min}=\mu$ is independent of $\beta$ and produces the vertical phase boundaries
of the metallic lobes. The $\alpha>0$ lobe retains these phase boundaries for
$0\leq\beta<1/4$, and the $\alpha<0$ lobe for $0\leq\beta<3/4$: 
\begin{align}
 1-\mu<\alpha<1+\mu,&\qquad 0\leq\beta<\frac14,
 \nonumber\\
 -1-\mu<\alpha<-1+\mu,&\qquad 0\leq\beta<\frac34 .
 \label{smeq:RM_mu_vertical_lobes}
\end{align}
Once this minimum passes through $x=L$, the endpoint is the last possible
metallic point. Its disappearance is set by $R(L)=\mu$, which traces the
curved phase boundaries, which we call caps, that close the lobes:
\begin{equation}
 \mu^2=1+\alpha^2-2\alpha\sin(2\pi\beta),
 \qquad
 \alpha_\pm(\beta)=\sin(2\pi\beta)
 \pm\sqrt{\mu^2-\cos^2(2\pi\beta)} .
 \label{smeq:RM_mu_cap_curves}
\end{equation}
The $\alpha>0$ curved phase boundary spans
\begin{equation}
 \frac14<\beta<\frac14+\frac{\arcsin\mu}{2\pi}
\end{equation}
and closes at $\alpha=\sqrt{1-\mu^2}$. The $\alpha<0$ curved phase boundary spans
\begin{equation}
 \frac34<\beta<\frac34+\frac{\arcsin\mu}{2\pi}
\end{equation}
and closes at $\alpha=-\sqrt{1-\mu^2}$.  The exact metallic width at
finite $\beta$ is
\begin{equation}
 w_\beta=\frac{L}{2\pi(1-\beta)}
 \left[\min\{\zeta^*+\phi_\mu,\,2\pi(1-\beta)\}
 -(\zeta^*-\phi_\mu)\right]_+,
 \qquad [q]_+=\max(q,0).
 \label{smeq:RM_mu_shortened_width}
\end{equation}
where we have used the definitions in \cref{smeq:RM_mu_angle}. Along the straight phase boundaries of the lobes, the local band edge varies
quadratically in space and is described by
\cref{smeq:RM_mu_quadratic_trap}, with $z=2$, $p=2$, and
$\vartheta=1/2$, so $\xi\sim L^{1/2}$ and $\Delta\sim L^{-1}$.  Along a
curved phase boundary, the metallic region reaches the open end with a nonzero
spatial slope. Let $s=L-x\geq0$ measure the real-space distance from the end and let
$r_b=R(L)-\mu$ be the endpoint band edge relative to the chemical
potential. It is positive on the insulating side and negative when the
endpoint is metallic. On a curved phase boundary $R(L)=\mu$. Away from its
junction with a straight phase boundary, $R'(L)<0$ on both caps. The
endpoint expansion and the resulting half-line theory are
\begin{equation}
 \begin{aligned}
 R(L-s)-\mu&=r_b+|g_L|s+\mathcal O(s^2/L^2),\\
 h_{\mathrm{edge}}&=-\frac{v_L^2\partial_s^2}{2\mu}+r_b+|g_L|s+\cdots .
 \end{aligned}
 \label{smeq:RM_mu_cap_theory}
\end{equation}
Here $v_L^2=1-\delta^2(L)=\sin^2(2\pi\beta)$.
On the curved phase boundary, the slope is
$|g_L|=|R'(L)|=[2\pi(1-\beta)/L]|\alpha\cos(2\pi\beta)|/\mu$.
Since $r_b=0$ and $g_L\sim L^{-1}$, we obtain
$(z,p,\vartheta)=(2,1,1/3)$,
$\xi_{\rm end}=[v_L^2/(2\mu|g_L|)]^{1/3}\sim L^{1/3}$, and
$\Delta\sim L^{-2/3}$.  Here ``straight'' and ``curved'' refer to the
shape of the phase boundary in the $(\alpha,\beta)$ diagram. The trap
order $p$ is determined instead by the dependence on the real-space
distance $X$ or $s$. The dots in \cref{fig:RM_unwinding}(b) mark where the spatial
minimum of $R(x)$ reaches $x=L$, so the straight and curved phase
boundaries meet. Consequently $R'(L)=g_L=0$: the linear term vanishes and the
leading spatial variation is quadratic,
restoring $(z,p,\vartheta)=(2,2,1/2)$.  Since the trapping minimum is then
at the end of the chain, the dot represents a crossover at the open end rather
than a generic point on the straight phase boundary. The vertical
and curved portions form one connected metal--insulator phase boundary,
yet have different trap orders, correlation lengths, and trap-level exponents.
This is the trap-scaling counterpart of multiversality
~\cite{AP_UC_Multiversality}.

\subsubsection{Boundary and defect-state crossings}

For $0<\beta<1$, the frozen Hamiltonians trace an open path in parameter
space. In a gapped texture sector, defining $\nu_\gamma$ requires
completing this path with a specified gapped interpolation from the
final Hamiltonian back to the initial one. The completion is part of
the boundary or seam data: the bulk profile alone does not fix the
loop invariant. Changing these boundary data can expel the trapped
charge through a localized crossing.

The defect crossing can be obtained explicitly for the open termination
used in the lattice model.  At $x=L$ the frozen endpoint parameters are
\begin{equation}
 \delta_L=\cos(2\pi\beta),\qquad
 J_L=\alpha-\sin(2\pi\beta).
 \label{smeq:RM_mu_endpoint}
\end{equation}
For $1/4<\beta<3/4$, $\delta_L<0$ and the right end supports the usual
Rice--Mele boundary state.  Writing
$t_{1,2}=(1\pm\delta_L)/2$, its wavefunction has support on the terminal
sublattice with successive amplitudes proportional to
$(-t_1/t_2)^j$~\cite{AP_ThoulessPhasediagrams}, and its energy is
\begin{equation}
 E_{\rm def}=J_L+\mathcal O(L^{-1}).
 \label{smeq:RM_mu_defect_energy}
\end{equation}
The state changes its occupation when it crosses the chemical potential,
which gives the locus
\begin{equation}
 \alpha_{\rm def}(\beta)=\sin(2\pi\beta)+\mu,
 \qquad \frac14<\beta<\frac34 .
 \label{smeq:RM_mu_defect_locus}
\end{equation}
With the definition in \cref{smeq:RM_local_Berry_phase}, the
occupied state is on the $\nu_\gamma=-1$ side of this line. At $\mu=0$
the line joins the two
terminations of the original trap-critical lines.  For $\mu>0$ it starts
at the outer phase boundary of the $\alpha>0$ metallic lobe and ends at the inner phase boundary
of the $\alpha<0$ lobe. This is the large-$L$ locus for the open
termination. Spatial gradients produce finite-size shifts.

\begin{figure}[H]
 \centering
 \begin{minipage}[t]{.66\linewidth}
  \vspace{0pt}\centering
  \includegraphics[width=\linewidth]{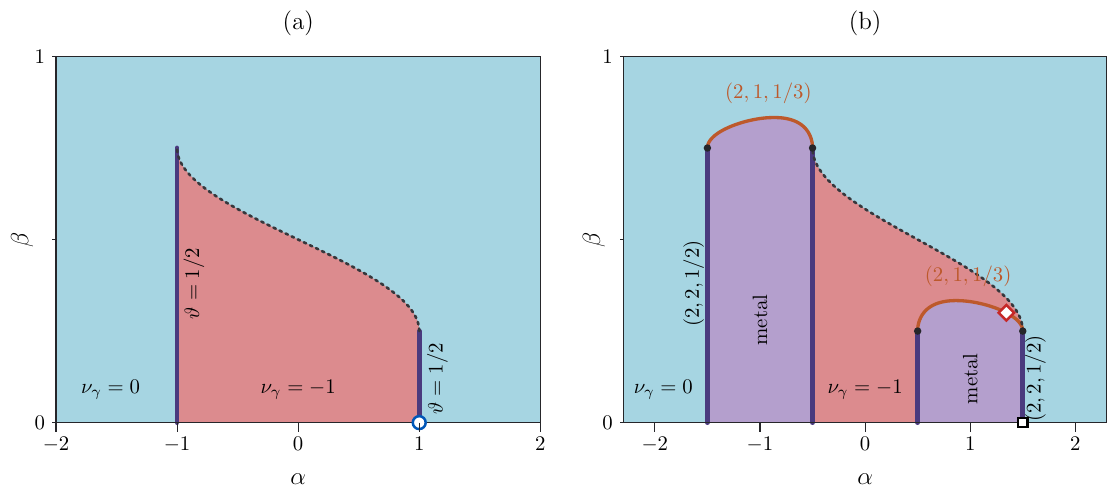}
 \end{minipage}\hfill
 \begin{minipage}[t]{.32\linewidth}
  \vspace{0pt}\centering
  \includegraphics[width=\linewidth]{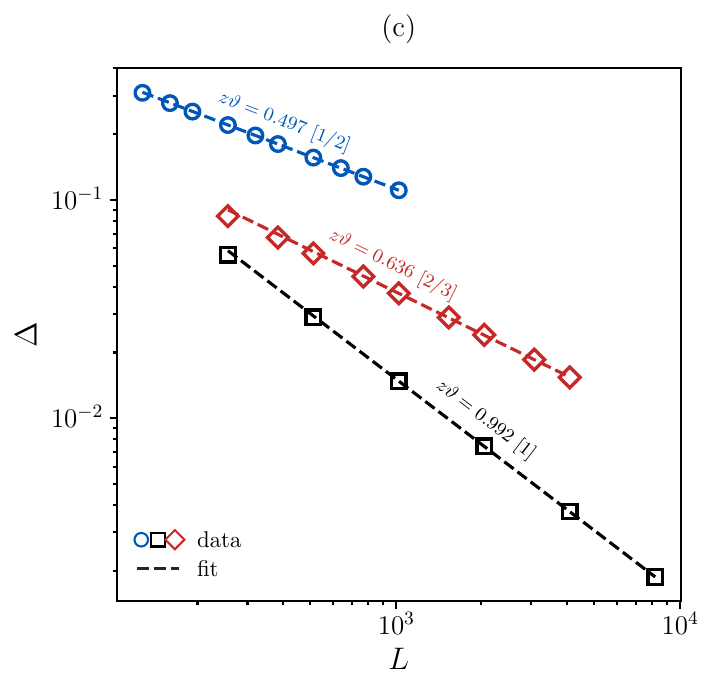}
 \end{minipage}
 \caption{Rice--Mele phase diagrams and trap-energy scaling.
 (a) At $\mu=0$, the trap-critical lines terminate.  (b) At $\mu=1/2$,
 they broaden into metallic lobes. The dotted curves in (a,b) are the
 open-termination defect-crossing loci. The filled dots in (b), at
 $(\alpha,\beta)=(1/2,1/4),(3/2,1/4),(-3/2,3/4),(-1/2,3/4)$,
 mark where the spatial minimum of $R(x)$ reaches the open endpoint. There the
 endpoint slope vanishes, and the $p=2$ theory on the straight phase boundary joins the
 $p=1$ theory on the curved phase boundary. The colored markers locate the parameters used in
 (c): the blue circle is the half-filled Dirac point
 $(\alpha,\beta;\mu)=(1,0;0)$, the black square is on the straight phase boundary at
 $(3/2,0;1/2)$, and the red diamond is on the curved phase boundary at
 $(1.344,0.30;1/2)$. (c) Exact-diagonalization data from the lattice
 Hamiltonian in \cref{smeq:RM_mu_Hamiltonian} for the corresponding
 adjacent low-energy spacings. At $\mu=0$, we omit the smaller
 finite-lattice displacement of the nominal zero mode when extracting
 the spacing. Dashed curves are fits to $\Delta=A L^b$:
 in the same order, $(A,b)=(3.46,-0.497),(14.37,-0.992)$, and
 $(3.06,-0.636)$. The labels on the fits give the extracted
 $z\vartheta=-b$, with the theoretical values in square brackets.}
 \label{fig:RM_unwinding}
\end{figure}

\Cref{fig:RM_unwinding}(a,b) summarizes the phase diagrams,
while panel (c) tests the predicted adjacent low-energy spacings:
$L^{-1/2}$ at half filling, $L^{-1}$ on a straight phase boundary,
and $L^{-2/3}$ on a curved phase boundary.

The exponents and leading scales are summarized in
\cref{tab:RM_mu_scaling}.

\begin{table}[H]
 \centering
 \begin{tabular}{lllll}
 Regime & $z$ & $p$ & $\vartheta$ & leading scales\\\hline
 Half-filled Dirac trap & 1 & 1 & $1/2$ & $\xi\sim L^{1/2}$, $\Delta\sim L^{-1/2}$\\
 Straight phase boundary (interior minimum) & 2 & 2 & $1/2$ & $\xi\sim L^{1/2}$, $\Delta\sim L^{-1}$\\
 Curved phase boundary (open endpoint) & 2 & 1 & $1/3$ & $\xi\sim L^{1/3}$, $\Delta\sim L^{-2/3}$\\
 Metallic interior & 1 & -- & -- & $w\sim L$, $\Delta\sim L^{-1}$
 \end{tabular}
 \caption{Scaling regimes of the Rice--Mele texture. $\Delta$ denotes
 the characteristic level spacing in each regime. The dynamical exponent in the last row is the linearized value at the two
 Fermi points. The extensive metal is not governed by a trap exponent.}
 \label{tab:RM_mu_scaling}
\end{table}

\subsection{Stability to weak perturbations and bosonization}
This subsection gives the field-theory details underlying the
stability statement in the main text.
To begin, let us specify the bosonization convention used in this work~\cite{AP_UC_Multiversality,AP_UC_Pump,AP_SEC_PhysRevB.108.245135}. We consider $\phi \equiv \phi + 2 \pi$ and $\theta \equiv \theta + 2 \pi$ to be canonically conjugate compact bosons with unit radius satisfying
	\begin{equation}
		[\partial_x \phi (x), \theta (x') ] = 2 \pi i \delta(x-x'),
		\label{eq:KacMoody}
	\end{equation}
We perturb the Rice--Mele chain by a nearest-neighbor interaction,
\begin{equation}
 \Ham=\Ham_{\rm RM}+U\sum_j
 \left(n_j-\frac12\right)\left(n_{j+1}-\frac12\right).
 \label{smeq:interacting_RM}
\end{equation}
Here $n_j=c_j^\dagger c_j$, and $U$ is measured in the hopping units used
in \cref{smeq:H_Rice_Mele}.
We bosonize the Dirac theory in \cref{smeq:Dirac_trapnode} near its
trap-critical points $(\alpha^*,x^*)$:
\begin{equation}
    \Ham \approx \frac{v}{2\pi} \int dx \left( \frac{1}{4K} (\partial_x \phi)^2 + K (\partial_x \theta)^2 \right)    - \mathcal{A}  \int dx ~\frac{(x-x^*)}{L} \cos\phi - \mathcal{B} (\alpha - \alpha^*) \int dx  \sin \phi. \label{smeq:H_Bosonize}
\end{equation}
The constants $\mathcal A$ and $\mathcal B$ are nonuniversal
mass amplitudes, while $v$ and $K$ are the renormalized velocity and
Luttinger parameter.
The two masses of the Dirac field $\psi$ in \cref{smeq:Dirac_trapnode} map to
\begin{equation}
 \psi^\dagger\sigma_1\psi\ \longleftrightarrow\ \cos\phi,
 \qquad
 \psi^\dagger\sigma_3\psi\ \longleftrightarrow\ \sin\phi .
 \label{smeq:Dirac_mass_bosonization}
\end{equation}
The microscopic charge conservation $\mathrm{U}(1)$ symmetry,
$c_x \mapsto e^{i\lambda} c_x$, acts as
$\theta \mapsto \theta + \lambda$. The Gaussian part of
\cref{smeq:H_Bosonize} is a $c=1$ conformal field theory, while
$\cos\phi$ and $\sin\phi$ both have scaling dimension $K$. The Luttinger
parameter $K$ and velocity $v$ depend on $U$ and can be obtained by Bethe
ansatz~\cite{Giamarchi,HALDANE_Bethe_LL_1981153}:
\begin{equation}
    K = \frac{\pi}{2 \arccos(-U)}, \quad v= \frac{K}{2K-1} \sin\left(\frac{\pi}{2K} \right)
\end{equation}
For $U=0$, $K=1$ and the theory is dual to the free Dirac Hamiltonian in \cref{smeq:Dirac_trapnode}.  At fixed chemical potential $\mu=0$ and for $-1/\sqrt{2}<U<1$, the two mass operators remain the only symmetry-allowed relevant perturbations: charge conservation forbids vertex operators of $\theta$, while higher harmonics of $\phi$ are irrelevant. Thus weak interactions cannot gap the trap-critical line, although they renormalize its location and scaling data. Since the modulated mass has RG eigenvalue $y_M=2-K$ and a simple zero $p=1$, the trap formula gives

\begin{equation}
 \vartheta=\frac{p}{p+y_M}=\frac{1}{3-K},
 \qquad \Delta\sim L^{-1/(3-K)} .
 \label{smeq:interacting_RM_theta}
\end{equation}
This reduces to $\vartheta=1/2$ at the free-fermion point and establishes stability throughout an open interval of interactions about it. In bosonization, the excess electric charge bound by the texture is given by the winding of $\moy{\phi(x)}$,
\begin{equation}
    :Q: = \frac{e}{2\pi} \int_{-\infty}^\infty dx~\moy{\partial_x \phi} \label{smeq:Q_bosonization}
\end{equation}
Here $:Q:$ is the electric charge, normal ordered relative to the homogeneous
half-filled background, and $e$ is the electric charge of an electron.
For $\alpha \neq \alpha^*$, the mass operators in
\cref{smeq:H_Bosonize} pin the field and gap the system.
At mean-field level, we obtain $\moy{\phi(x)}$ by minimizing the local
potential density,
\begin{equation}
 V_{\rm loc}(x,\phi)=-\mathcal A\frac{x-x^*}{L}\cos\phi
 -\mathcal B(\alpha-\alpha^*)\sin\phi.
\end{equation}
Across $\alpha=\alpha^*$, the winding of $\moy{\phi(x)}$ changes by one. Equation~\eqref{smeq:Q_bosonization} then gives $|Q_{\alpha<\alpha^*} - Q_{\alpha>\alpha^*}| = |e|$, recovering the result in the main text.

\section{Two-dimensional class D Kitaev-chain texture: additional details}
\label{app:classD_texture}

\subsection{Homogeneous family and occupied-band invariant}
This section contains additional details of the Class-D texture that can trap Kitaev's Majorana chain discussed in the main text. We begin with a stack of Kitaev chains~\cite{KitaevMajoranaChain}
running along $x$. The coordinate
$x=0,\ldots,L_x-1$ labels sites along a chain, while $y=0,\ldots,2L_y-1$
labels the layers. Because the interlayer coupling alternates, we group
layers $y=2j$ and $2j+1$ into a unit cell labeled by
$j=0,\ldots,L_y-1$. No physical many-body symmetry is imposed: fermion
parity is required by locality. Each site carries two self-adjoint
Majorana operators, $a_{x,y}^\dagger=a_{x,y}$ and
$b_{x,y}^\dagger=b_{x,y}$, which obey
\begin{equation}
 \{a_{x,y},a_{x',y'}\}=\{b_{x,y},b_{x',y'}\}
 =2\delta_{xx'}\delta_{yy'},\qquad
 \{a_{x,y},b_{x',y'}\}=0.
\end{equation}
With the microscopic energy scale set to unity, write
$\Ham=\Ham_0+\Ham_\rho$, where
\begin{align}
 \Ham_0(m,\delta)&=\frac{i}{4}\sum_{x,y}\Big[
 (1+m)a_{x,y}b_{x,y}+(1-m)b_{x,y}a_{x+1,y}+\big(1+(-1)^y\delta\big)
 (a_{x,y}a_{x,y+1}-b_{x,y}b_{x,y+1})\Big],
 \label{smeq:classD_H0}\\
 \Ham_\rho(\rho)&=\frac{i\rho}{4}\sum_{x,y}
 (a_{x,y}b_{x,y+1}-b_{x,y}a_{x,y+1}).
 \label{smeq:classD_Hrho}
\end{align}
The first two terms of $\Ham_0$ couple Majoranas within each chain:
$m$ tunes the chain between the atomic and Kitaev limits. The last term
couples neighboring layers, with $\delta$ controlling the alternating
bond strengths. The additional interlayer coupling $\Ham_\rho$ splits
the two Majorana masses at the diabolical point. We return to its effect later. Let us first set $\rho=0$.

The $\mathbb Z_2$ classification of the Kitaev chain allows two chains to
be coupled and deformed into a trivial gapped state. Thus a Kitaev chain
is its own inverse under stacking. In $\Ham_0$, there are two ways to
pair the layers: $\delta=+1$ couples $(2j,2j+1)$, while $\delta=-1$
couples $(2j-1,2j)$. Varying $m$ and $\delta$ along a loop around
$m=\delta=0$ interpolates between these two ways of trivializing pairs of
chains. This is the first suspension of the Kitaev chain. If the layer
pairing terminates, one Kitaev chain remains unpaired, exposing  the
lower-dimensional phase trapped by the spatial texture
~\cite{KitaevMajoranaChain,AP_ThoulessPhasediagrams,Wenetal_topologicalfamilies}.

\begin{figure}[tbp]
 \centering
 \includegraphics[width=\linewidth]{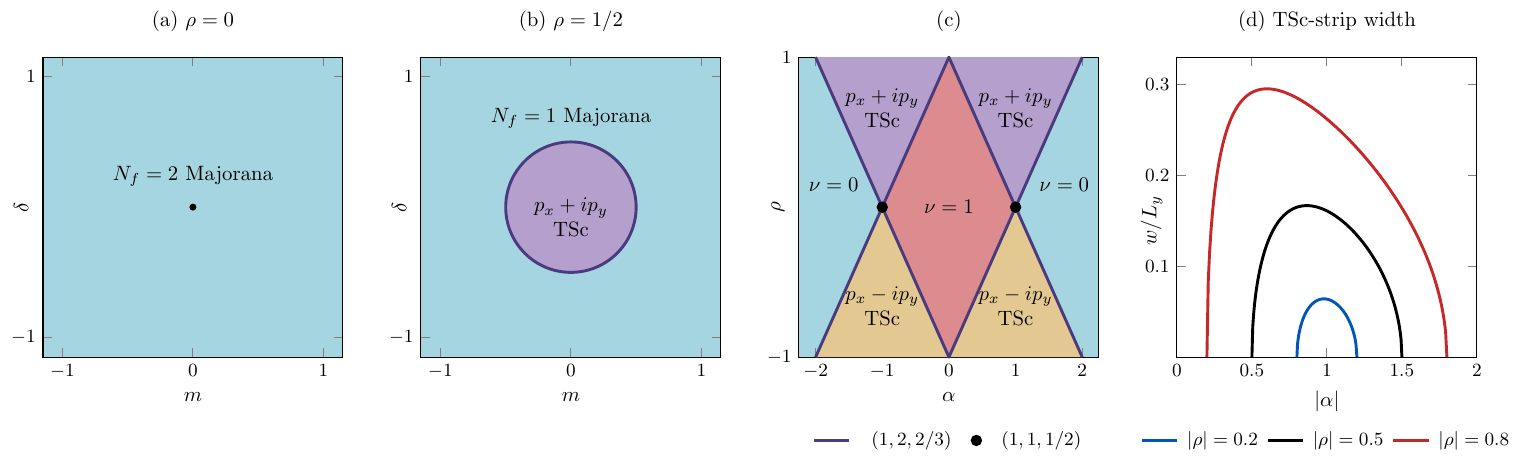}
 \caption{Homogeneous class D phase diagrams and the strip
 produced by a spatial texture.
 (a) At $\rho=0$, the homogeneous system is gapped with $C=0$ except at
 $m=\delta=0$, as follows from \cref{smeq:classD_MD0_spectrum}.
 The conformal field theory at $m=\delta=0$ contains $N_f=2$ relativistic
 Majorana fermions~\cite{ReadGreen2000,Ryu}. (b) At $\rho=1/2$, the
 $N_f=1$ critical surface $m^2+\delta^2=\rho^2$ appears as a circle: its
 exterior has $C=0$, while its interior is a $p_x+i p_y$ topological
 superconductor (TSc) with $C=1$,
 as given by \cref{smeq:classD_rho_lattice_spectrum,smeq:classD_Chern}.
 For $\rho<0$, the interior is $p_x-i p_y$ with $C=-1$.
 (c) Phase diagram of the spatial texture in
 \cref{smeq:classD_spatial_texture} in the $(\alpha,\rho)$ plane for
 $|\rho|<1$. The straight phase boundaries $|\alpha|=1\pm|\rho|$ separate the
 gapped $\nu=1$ and $\nu=0$ sectors from extensive TSc strips, as in
 \cref{smeq:classD_alpha_phases}. The sign of $\rho$ selects the strip
 chirality. Legend entries give $(z,p,\vartheta)$: the lines are the
 quadratic Majorana traps of \cref{smeq:classD_tangent_exponent}, and the
 black points $(\alpha,\rho)=(\pm1,0)$ are the linear two-Majorana traps
 of \cref{smeq:classD_MD0_scaling}.
 (d) The TSc-strip fraction $w/L_y$ for the spatial texture in
 \cref{smeq:classD_spatial_texture}, evaluated from
 \cref{smeq:classD_droplet_width}, is nonzero for $1-|\rho|<|\alpha|<1+|\rho|$, giving an extensive area
 $L_xw$.}
 \label{fig:classD_homogeneous}
\end{figure}

For spatially uniform $m$ and $\delta$, we use translation invariance to
obtain the bulk spectrum and occupied states. With periodic boundaries,
the Fourier-transformed Majorana spinor in a two-layer cell is
\begin{equation}
 \Psi_{\vec{k}}=\frac{1}{\sqrt{2L_xL_y}}
 \sum_{x=0}^{L_x-1}\sum_{j=0}^{L_y-1}
 e^{-i(k_xx+k_yj)}
 \begin{pmatrix}a_{x,2j}&b_{x,2j}&a_{x,2j+1}&b_{x,2j+1}\end{pmatrix}^{\mathsf T},
 \qquad \Psi_{\vec{k}}=\bigl[\Psi_{-\vec{k}}^{\dagger}\bigr]^{\mathsf T}.
 \label{smeq:classD_Fourier}
\end{equation}
Here $k_x$ is conjugate to the site coordinate $x$ and $k_y$ to the cell
coordinate $j$. The self-conjugacy condition follows from the Hermiticity
of the real-space Majorana operators. Substitution in
\cref{smeq:classD_H0} gives the Bogoliubov--de Gennes (BdG) Hamiltonian
\begin{align}
 \Ham_0&=\frac12\sum_{\vec{k}}\Psi_{\vec{k}}^{\dagger}
 h_0(\vec{k};m,\delta)\Psi_{\vec{k}},\nonumber\\
 h_0(\vec{k};m,\delta)
 &=\frac{1-m}{2}\sin k_x\,\Gamma_1
  +\frac{1-\delta}{2}\sin k_y\,\Gamma_2+\left(\frac{1+m}{2}-\frac{1-m}{2}\cos k_x\right)\Gamma_3
  +\left(\frac{1+\delta}{2}-\frac{1-\delta}{2}\cos k_y\right)\Gamma_4.
 \label{smeq:classD_first_quantized}
\end{align}
The coefficients of $\Gamma_a$ in \cref{smeq:classD_first_quantized}
define $\vec{d}=(d_1,d_2,d_3,d_4)$, so $h_0=\vec{d}\cdot\vec{\Gamma}$.
We use the representation
\begin{equation}
 \Gamma_1=-\mathbb1\otimes\sigma_x,\quad
 \Gamma_2=-\sigma_x\otimes\sigma_z,\quad
 \Gamma_3=-\mathbb1\otimes\sigma_y,\quad
 \Gamma_4=-\sigma_y\otimes\sigma_z,
 \label{smeq:classD_gammas}
\end{equation}
where $\{\Gamma_a,\Gamma_b\}=2\delta_{ab}$. The matrices $\Gamma_1,\Gamma_2$ are real and multiply
odd functions of momentum. $\Gamma_3,\Gamma_4$ are imaginary and multiply
even functions. Consequently,
\begin{equation}
 \mathcal C h_0(\vec{k})\mathcal C^{-1}=h_0(\vec{k})^*
 =-h_0(-\vec{k}),\qquad \mathcal C=K,\qquad \mathcal C^2=1,
 \label{smeq:classD_PHS}
\end{equation}
where $K$ denotes complex conjugation. This particle--hole constraint on
the first-quantized BdG Hamiltonian expresses the Majorana redundancy in
\cref{smeq:classD_Fourier}. At the many-body level, locality requires
fermion parity to be preserved, with no additional symmetry imposed
~\cite{kitaevclassification,SchnyderRyu,Ryu}.
Anticommutation of the $\Gamma_a$ makes $h_0^2$ proportional to the
identity, with coefficient
\begin{equation}
 |\vec{d}|^2=\sin^2\frac{k_x}{2}+m^2\cos^2\frac{k_x}{2}
 +\sin^2\frac{k_y}{2}+\delta^2\cos^2\frac{k_y}{2}.
 \label{smeq:classD_MD0_spectrum}
\end{equation}
The energies are therefore $\pm|\vec{d}|$, each with twofold
degeneracy. The only gap closing is at $\vec{k}=0$, $m=\delta=0$
[\cref{fig:classD_homogeneous}(a)].
Every other point in this parameter plane has the same two-dimensional
strong invariant, $C=0$, where $C$ is the first Chern number of the
occupied, negative-energy BdG bands~\cite{ReadGreen2000,Ryu}. This corresponds to the topologically trivial $p-$wave superconductor.

\subsubsection{Occupied-band invariant}
The Chern number characterizes each fixed Hamiltonian. To characterize
a family, we must also follow how its occupied states change as the
parameters vary. Consider a closed path $(m(\lambda),\delta(\lambda))$,
with $0\leq\lambda<2\pi$, along which the Hamiltonian remains gapped.
A path that winds once around the gap closing at $m=\delta=0$ gives a
nontrivial family, even though every member has $C=0$. It cannot be
shrunk to a single Hamiltonian through gapped class D families. A loop that does not enclose $m=\delta=0$ can be contracted and is trivial. The
$\mathbb Z_2$ invariant $\nu$ distinguishes these two cases
~\cite{TeoKane2010TopologicalDefects}.

The calculation parallels the Berry-phase calculation for Rice--Mele,
but now involves two momentum coordinates and the loop parameter.
Together they form $X=T^2_{\mathrm{BZ}}\times S^1_\lambda$: the
two-dimensional Brillouin zone times the parameter circle. Define
$\vec{\xi}=(k_x,k_y,\lambda)$,
$\partial_\mu=\partial/\partial\xi^\mu$, and the non-Abelian Berry
connection~\cite{WilczekZee1984}
\begin{equation}
 A^\mu_{mn}=i\langle u_m|\partial_\mu u_n\rangle.
\end{equation}
Here $|u_n(\vec{\xi})\rangle$ are occupied BdG eigenstates, and
$m,n$ are band indices. The class D family invariant is
the Chern--Simons integral~\cite{TeoKane2010TopologicalDefects}
\begin{equation}
 \nu=\frac{1}{4\pi^2}\int_Xd^3\xi\,
 \epsilon_{\mu\nu\sigma}\operatorname{Tr}\!\left(
 A^\mu\partial_\nu A^\sigma-\frac{2i}{3}A^\mu A^\nu A^\sigma\right)
 \bmod2.
 \label{smeq:classD_CS_invariant}
\end{equation}
The trace is over occupied bands, the antisymmetric symbol satisfies
$\epsilon_{123}=1$, and repeated coordinate indices are summed.
The BdG particle--hole constraint in \cref{smeq:classD_PHS} quantizes this
expression to an integer. A smooth change of basis among occupied states
changes it by an even integer, leaving its parity well defined~\cite{TeoKane2010TopologicalDefects}.

For the four-band Hamiltonian at $\rho=0$, we can also evaluate the invariant
using $\hat{\vec{d}}=\vec{d}/|\vec{d}|$, with $\vec{d}$ formed from the
coefficients in \cref{smeq:classD_first_quantized}. Its winding-number
representation~\cite{Ryu} is
\begin{equation}
 W_3=\frac{1}{2\pi^2}\int dk_x\,dk_y\,d\lambda\,
 \epsilon^{abcd}\hat d_a\partial_{k_x}\hat d_b
 \partial_{k_y}\hat d_c\partial_\lambda\hat d_d,
 \qquad \nu=W_3\bmod2,
 \label{smeq:classD_W3}
\end{equation}
$W_3$ counts, with orientation, how many times $\hat{\vec{d}}$ covers the
unit three-sphere as $k_x,k_y,\lambda$ vary. 

\subsubsection{Homogeneous phase diagram at nonzero $\rho$}

The $\rho=0$ spectrum has a single gap closing diabolical point in the parameter plane.
We now include the
fermion-parity-even coupling $\Ham_\rho$ in \cref{smeq:classD_Hrho}
and determine how it changes this homogeneous phase diagram. The
coupling splits the two Majorana masses that vanish together at
$m=\delta=0$. In the Bloch basis of
\cref{smeq:classD_first_quantized}, the coupling becomes
\begin{equation}
 h_\rho(\vec{k})=-\frac{i\rho}{2}\Gamma_1
 \big[(1+\cos k_y)\Gamma_2+\sin k_y\Gamma_4\big].
 \label{smeq:classD_rho_Bloch}
\end{equation}

Near the diabolical point, let $\vec{q}=(q_x,q_y)$ denote momentum
measured from $\vec{k}=0$. The continuum Hamiltonian is
\begin{equation}
 h_{\mathrm{continuum}}(\vec{q})=\frac{q_x}{2}\Gamma_1+\frac{q_y}{2}\Gamma_2
 +m\Gamma_3+\delta\Gamma_4+\rho\Gamma_D,
 \qquad \Gamma_D=-i\Gamma_1\Gamma_2.
 \label{smeq:classD_continuum_Hamiltonian}
\end{equation}
The matrix $\Gamma_D$ anticommutes with the kinetic matrices but
commutes with the original mass matrices:
\begin{equation}
 \{\Gamma_D,\Gamma_{1,2}\}=0,\qquad
 [\Gamma_D,\Gamma_{3,4}]=0.
 \label{smeq:classD_MD_algebra}
\end{equation}
The combined original mass has magnitude $R=\sqrt{m^2+\delta^2}$,
the distance from the diabolical point in parameter space, and matrix
$\Gamma_R=(m\Gamma_3+\delta\Gamma_4)/R$ for $R>0$.
Since $Q_R=\Gamma_D\Gamma_R$ commutes with the homogeneous Hamiltonian,
we can separate the four-component problem into two two-component
Majorana sectors labeled by its eigenvalue $q_R=\pm1$. Within each
sector, the mass is $(R+q_R\rho)\Gamma_R$: the perturbation increases
one mass and decreases the other. The two dispersions are therefore
\begin{equation}
 E_\pm^2(\vec{q})=\frac{q_x^2+q_y^2}{4}+(\rho\pm R)^2.
 \label{smeq:classD_continuum_spectrum}
\end{equation}
To determine the range over which this continuum description captures
all gap closings, we also diagonalize the full Bloch Hamiltonian
$h_0+h_\rho$. Its spectrum is
\begin{equation}
 E_\pm^2(\vec{k})=d_1^2+\sin^2\frac{k_y}{2}
 +\left(\sqrt{d_3^2+\delta^2\cos^2\frac{k_y}{2}}
       \pm\rho\cos\frac{k_y}{2}\right)^2.
 \label{smeq:classD_rho_lattice_spectrum}
\end{equation}
Here $d_1$ and $d_3$ are the coefficients in \cref{smeq:classD_first_quantized}.
For $|\rho|<1$, zero energy requires $k_x=k_y=0$ and $R=|\rho|$.
The gap closes on $m^2+\delta^2=\rho^2$, as predicted by
\cref{smeq:classD_continuum_spectrum}. For $|\rho|<1$, the full lattice
spectrum confirms that this is the only gap-closing phase boundary. On this circle,
one Majorana is massless and the other has mass $2|\rho|$.
For $|\rho|\geq1$, \cref{smeq:classD_rho_lattice_spectrum} develops
additional lattice gap closings, so the single-node continuum theory and
two-lobe phase diagram below are no longer complete. We do not consider
this regime further.
For the finite-$\rho$ analysis below we take $0<|\rho|<1$. Expanding the
branch that becomes gapless at a point $(m_*,\delta_*)$ on the circle
gives $v_x=(1-m_*)/2$ and $v_y=1/2$. We rescale the continuum coordinate
$x$ by a constant to set $v_x=1/2$. Transverse positions and droplet
widths retain their original cell units. The circle separates phases
with different Chern numbers of the occupied BdG bands:
\begin{equation}
 C=\frac12\left[\operatorname{sgn}(\rho+R)
 +\operatorname{sgn}(\rho-R)\right]
 =\begin{cases}\operatorname{sgn}\rho,&R<|\rho|,\\0,&R>|\rho|.
 \end{cases}
 \label{smeq:classD_Chern}
\end{equation}
The interior is a $p_x+i p_y$ TSc phase for $\rho>0$ and a
$p_x-i p_y$ TSc phase for $\rho<0$
~\cite{ReadGreen2000,TeoKane2010TopologicalDefects}.
\Cref{fig:classD_homogeneous}(a,b) compares these homogeneous phase
diagrams: the isolated closing at $\rho=0$ expands into a critical circle
at $\rho=1/2$. Both panels refer to spatially uniform couplings. The
curves have the same shapes as in \cref{fig:RM_homogeneous_droplet}
because their phase boundaries obey the same equations. In Rice--Mele,
the local band edge $R_{\rm RM}=\sqrt{J^2+\delta^2}$ reaches the Fermi
level at $R_{\rm RM}=|\mu|$. In class D, the lighter Majorana mass
vanishes at $R_{\rm D}=\sqrt{m^2+\delta^2}=|\rho|$. Both homogeneous
phase boundaries are therefore circles. For the sinusoidal spatial textures,
both radial profiles satisfy $R^2=1+\alpha^2+2\alpha\sin\zeta$, where
$\zeta$ is the spatially varying angle along the parameter loop.
Replacing $|\mu|$ by $|\rho|$ gives the same lobe phase boundaries and the
same fraction of the sample with $R$ below the threshold, explaining
the identical droplet-width curves. The interiors have different
physics: a metal in Rice--Mele and a gapped TSc phase in class D.

\FloatBarrier
\subsection{Spatial texture and the linear trap}
So far, $\lambda$ has labeled spatially uniform Hamiltonians. We now
embed the family in a single sample by letting $\lambda$ vary slowly
across the layers, while preserving translation invariance along the
chains. For the continuum profile, we measure $y$ and the transverse
size $L_y$ in two-layer cell units, since $k_y$ is conjugate to the
cell coordinate. Changing from layer to cell units rescales both
lengths and leaves $y/L_y$ unchanged. We first set $\rho=0$ and choose
$\lambda(y)=2\pi y/L_y$. A uniform offset $\alpha$ shifts the center of
the parameter loop:
\begin{equation}
 m(y)=\alpha+\sin\frac{2\pi y}{L_y},\qquad
 \delta(y)=\cos\frac{2\pi y}{L_y}.
 \label{smeq:classD_spatial_texture}
\end{equation}
As $y$ runs across the sample, $(m(y),\delta(y))$ traces a unit circle
centered at $(\alpha,0)$. For $|\alpha|<1$ the circle encloses the gap
closing and $\nu=1$. For $|\alpha|>1$ it does not, and $\nu=0$. The
nontrivial texture carries a Kitaev phase along the $x$ direction
~\cite{KitaevMajoranaChain,Wenetal_topologicalfamilies}.
At the transition, the parameter circle passes through $m=\delta=0$, so
the local homogeneous gap vanishes at one transverse position. We obtain
the theory near this position by expanding the Bloch Hamiltonian about
$\vec{k}=0$ and replacing the transverse momentum by $-i\partial_y$.
The velocities along and across the chains are both $1/2$ at the node. Finite changes
away from the node can be absorbed into constant coordinate rescalings
and do not change the trap exponents.
At $\alpha^*=1$ the loop reaches the diabolical point at $y_*/L_y=3/4$.
At $\alpha^*=-1$ it does so at $y_*/L_y=1/4$. We write
\begin{equation}
 t=\alpha-\alpha^*,\qquad
 \ell_y=\sqrt{\frac{L_y}{2\pi}},\qquad
 u=\frac{y-y_*}{\ell_y^2},\qquad s_*=\alpha^*=\pm1.
 \label{smeq:classD_kappa}
\end{equation}
Expanding the two masses gives
\begin{equation}
 m=t+\frac{s_*u^2}{2}+\mathcal O(u^4),\qquad
 \delta=s_*u+\mathcal O(u^3).
 \label{smeq:classD_MD0_masses}
\end{equation}
Keeping the leading terms in momenta and $u$ yields
\begin{align}
 \Ham_0&\approx\frac12\int d^2r\,
 \Psi^\dagger(\vec{r})h\Psi(\vec{r}),\nonumber\\
 h&=-\frac{i}{2}\Gamma_1\partial_x
 -\frac{i}{2}\Gamma_2\partial_y+t\Gamma_3
 +s_*\frac{y-y_*}{\ell_y^2}\Gamma_4.
 \label{smeq:classD_MD0_trap_Hamiltonian}
\end{align}
The matrices $\Gamma_a$ are given in \cref{smeq:classD_gammas}.
The four-component field is self-conjugate,
$\Psi_{\vec{k}}=[\Psi_{-\vec{k}}^\dagger]^T$ and
$\Psi(\vec{r})=[\Psi^\dagger(\vec{r})]^T$: the prefactor $1/2$
removes BdG double counting. The four real components describe two
relativistic Majorana species in two dimensions, equivalently one
complex Dirac fermion~\cite{ReadGreen2000,Ryu}. The term $t\Gamma_3$ tunes through the transition,
while the mass multiplying $\Gamma_4$ grows linearly with the spatial displacement
$y-y_*$ and confines low-energy states near $y_*$. Writing $g=s_*/\ell_y^2$ and $v_x=v_y=1/2$, the square of the Hamiltonian
at fixed $k_x$ is
\begin{equation}
 h^2=v_x^2k_x^2+t^2-v_y^2\partial_y^2
 +g^2(y-y_*)^2-v_yg\,S,\qquad S=i\Gamma_2\Gamma_4,\quad S^2=1.
 \label{smeq:classD_linear_square}
\end{equation}
The $y$-dependent part is a harmonic oscillator, as in the Rice--Mele
trap. In the sector $S=\operatorname{sgn}g$, the constant matrix term
cancels the oscillator zero-point energy. The resulting transverse
ground state has a Gaussian envelope. Projection onto this state leaves
a two-component one-dimensional Majorana Hamiltonian along $x$, with
dispersion $E_0(k_x)=\pm\sqrt{v_x^2k_x^2+t^2}$. Its mass changes sign
through $t=0$, where the gap closes and the one-dimensional class D
$\mathbb Z_2$ invariant changes. The higher transverse levels satisfy
$E_n^2=v_x^2k_x^2+t^2+2n v_y|g|$, $n\geq1$.
The linear mass profile has $p=1$ and $z=1$, so
\cref{smeq:trap_scaling} gives $\vartheta_0=1/2$.
Their transverse length and spacing are
\begin{equation}
 \ell_0=\sqrt{\frac{v_y}{|g|}}=\frac{\ell_y}{\sqrt2}
 \propto L_y^{1/2},\qquad
 \Delta_\perp\sim\frac{v_y}{\ell_0}\propto L_y^{-1/2},
 \qquad \vartheta_0=\frac12.
 \label{smeq:classD_MD0_scaling}
\end{equation}
The critical region extends along the full chain direction and has area
$\mathcal O(L_xL_y^{1/2})$. Its area grows with system size while its
fraction of the sample vanishes, as expected for trap-scaling criticality
~\cite{CampostriniVicari2009PRL_TSS,CampostriniVicari2010PRA_QTSS}.
On this transverse scale, the quadratic correction to $m$ in
\cref{smeq:classD_MD0_masses} is subleading. The energy
$\Delta_\perp$ measures the spacing between transverse excitations. The
lowest Majorana branch has a separate finite-size scale set by $L_x$.

\FloatBarrier
\subsection{TSc strips, interface modes, and phase boundaries}

\subsubsection{Extensive droplets: phase boundaries and spatial size}
  With the same spatial profile in \cref{smeq:classD_spatial_texture},
each transverse position selects a
point on the parameter circle. Wherever this point lies inside the
homogeneous TSc disk, the corresponding layers form part of a TSc
strip. The distance of $(m(y),\delta(y))$ from $m=\delta=0$ in parameter space is
\begin{equation}
 R^2(y)=1+\alpha^2+2\alpha\sin\frac{2\pi y}{L_y},\qquad
 R_{\min}=\bigl||\alpha|-1\bigr|.
 \label{smeq:classD_R_texture}
\end{equation}
In the slowly varying limit, this local bulk description applies away
from the interfaces. The mass of the Majorana species that becomes
gapless at an interface is $\mu_L(y)=R(y)-|\rho|$: we refer to it as the
light mass. It is negative in the TSc region. Comparing $R_{\min}$
with $|\rho|$ gives three regimes:
\begin{equation}
 \begin{array}{ccl}
 |\alpha|<1-|\rho|&:&\nu=1\text{ gapped Kitaev texture},\\
 1-|\rho|<|\alpha|<1+|\rho|&:&p_x\pm i p_y\text{ strip},\\
 |\alpha|>1+|\rho|&:&\nu=0\text{ gapped texture}.
 \end{array}
 \label{smeq:classD_alpha_phases}
\end{equation}
In either gapped regime, $\rho$
can be reduced to zero without any local Hamiltonian becoming gapless,
so the $\nu$ labels agree with those of the original Kitaev texture.
\Cref{fig:classD_homogeneous}(c) shows these sectors in the
$(\alpha,\rho)$ plane. The intervening TSc strips separate the two
gapped texture sectors, with phase boundaries $|\alpha|=1\pm|\rho|$.

\begin{figure}[t]
 \centering
 \includegraphics[width=\linewidth]{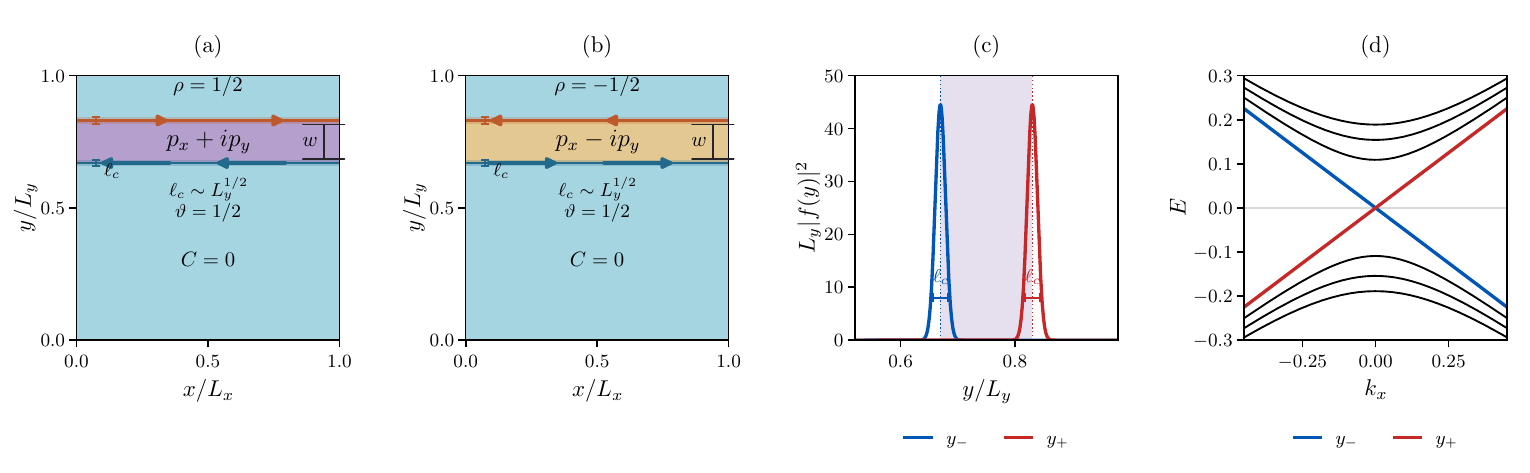}
 \caption{Extensive TSc strips and their chiral interface
 modes at $\alpha=1$.
 (a,b) Schematics for $\rho=1/2$ and $-1/2$, respectively. The interface
 positions $y_\pm$ and strip width
 $w/L_y=2\arcsin(1/4)/\pi$ follow from
 \cref{smeq:classD_droplet_width}. The narrow shaded bands around
 $y_\pm$ show the Majorana localization length
 $\ell_c=\sqrt{v_y/|g_c|}\propto L_y^{1/2}$ from
 \cref{smeq:classD_interface_gradient,smeq:classD_interface_width}: hence
 $\vartheta=1/2$, the transverse spacing is $\mathcal O(L_y^{-1/2})$,
 and $w/\ell_c\to\infty$. The arrows are the propagation
 directions fixed by \cref{smeq:classD_edge_dispersion}.
 The horizontal direction is periodic.
 (c) The two Gaussian densities follow directly from
 \cref{smeq:classD_interface_width} for $\rho=1/2$ and $L_y=512$:
 $y_\pm/L_y\simeq0.670,0.830$ and $\ell_c\simeq6.49$ two-layer cells.
 (d) The colored chiral branches follow from
 \cref{smeq:classD_edge_dispersion}. The black curves are the first three
 massive transverse levels from \cref{smeq:classD_interface_ladder}.
 All four panels use the continuum expressions.}
 \label{fig:classD_droplets}
\end{figure}

To find the strip interfaces, write $\theta=2\pi y/L_y$. Let $\theta_*$
be the angle at which the loop comes closest to $m=\delta=0$, and let
$\phi_0$ be the angular half-width of the portion inside the TSc disk:
\begin{equation}
 \theta_*=
 \begin{cases}3\pi/2,&\alpha>0,\\\pi/2,&\alpha<0,\end{cases}
 \qquad
 \phi_0=\arccos\!\left(\frac{1+\alpha^2-\rho^2}{2|\alpha|}\right).
 \label{smeq:classD_droplet_angle}
\end{equation}
Using $\theta_*$ and $\phi_0$ from \cref{smeq:classD_droplet_angle}, inside a lobe we have
$R^2=(|\alpha|-1)^2+
2|\alpha|[1-\cos(\theta-\theta_*)]$, so the TSc interval is
$\theta_*-\phi_0<\theta<\theta_*+\phi_0$. Its interface positions and width are
\begin{equation}
 y_\pm=\frac{L_y}{2\pi}(\theta_*\pm\phi_0),\qquad
 w=y_+-y_-=
 \frac{L_y}{\pi}\arccos\!\left(
 \frac{1+\alpha^2-\rho^2}{2|\alpha|}\right).
 \label{smeq:classD_droplet_width}
\end{equation}
\Cref{fig:classD_homogeneous}(d) plots this width as a fraction of $L_y$
for several values of $|\rho|$.
There is no TSc interval outside the lobe. For fixed $0<|\rho|<1$ and $\alpha$
in its interior, $w/L_y$ is a nonzero constant. The droplet is therefore
a strip of extensive area $L_xw=\mathcal O(L_xL_y)$. For fixed $|\rho|$, the maximum occurs at
$|\alpha|=\sqrt{1-\rho^2}$ and equals
$w_{\max}/L_y=\arcsin|\rho|/\pi$.

At fixed $|\rho|$, the two strip interfaces merge as $|\alpha|$ reaches
either phase boundary $1\pm|\rho|$. The macroscopic width vanishes.
A critical region of width
$\mathcal O(L_y^{2/3})$ remains at the transition. We derive this scale
after discussing the separated interfaces inside the lobe.

\subsubsection{Chiral Majorana modes at the strip interfaces}
Inside a lobe, the TSc strip is bounded by two interfaces with the
surrounding $C=0$ region. The Chern number changes by one across each
interface, so each supports a chiral Majorana mode
~\cite{ReadGreen2000,TeoKane2010TopologicalDefects}.

\Cref{fig:classD_droplets}(a,b) shows the strip and its two interface
channels.
We obtain each mode's wavefunction and dispersion by solving the
Majorana mass-domain-wall problem
~\cite{JackiwRebbi1976,ReadGreen2000}. Taking $\rho>0$,
$\mu_L(y)=R(y)-\rho$ changes sign at each interface $y_c=y_\pm$.
Its leading expansion there is
\begin{equation}
 \mu_L(y)=g_c(y-y_c)+\cdots,\qquad
 |g_c|=\frac{2\pi}{L_y}\frac{|\alpha\cos\theta_c|}{\rho},\qquad
 \sin\theta_c=\frac{\rho^2-\alpha^2-1}{2\alpha}.
 \label{smeq:classD_interface_gradient}
\end{equation}
Here $\theta_c=2\pi y_c/L_y$ and $g_c=\partial_y\mu_L(y_c)$ is the
mass gradient, negative at $y_-$ and positive at $y_+$. In a
two-component basis for the light Majorana, the interface Hamiltonian is
\begin{equation}
 h_{\mathrm{interface}}(k_x)=v_x k_x\sigma_x-i v_y\sigma_z\partial_y
 +g_c(y-y_c)\sigma_y,\qquad v_x=v_y=\frac12.
 \label{smeq:classD_interface_theory}
\end{equation}
At $k_x=0$, multiplying the zero-mode equation by $i\sigma_z$ gives
$v_y\partial_y f+g_c(y-y_c)\sigma_xf=0$. The normalized solution is
\begin{equation}
 f_c(y)=\frac{e^{-(y-y_c)^2/(2\ell_c^2)}}{(\pi\ell_c^2)^{1/4}}\,u_c,
 \qquad \sigma_xu_c=\operatorname{sgn}(g_c)u_c,
 \qquad \ell_c=\sqrt{\frac{v_y}{|g_c|}}\propto L_y^{1/2}.
 \label{smeq:classD_interface_width}
\end{equation}
Here $u_c$ is the normalized, $y$-independent eigenvector of $\sigma_x$
with eigenvalue $\operatorname{sgn}(g_c)$. The two Gaussian densities in
\cref{fig:classD_droplets}(c) are $|f_c(y)|^2$ at $y_-$ and $y_+$.
Their separation is the strip width $w$.
Projecting $h_{\mathrm{interface}}$ onto this transverse state gives its
dispersion along the interface:
\begin{equation}
 E_c(k_x)=\operatorname{sgn}(g_c)v_xk_x.
 \label{smeq:classD_edge_dispersion}
\end{equation}
There is one chiral Majorana channel at each interface, with opposite
propagation directions on the two sides of the strip. Reversing $\rho$
reverses both directions, as required by bulk--boundary correspondence for
the jump of the local Chern number
~\cite{ReadGreen2000,TeoKane2010TopologicalDefects}.
The higher transverse levels obey
\begin{equation}
 E_{n,\pm}(k_x)=\pm\sqrt{v_x^2k_x^2+2n v_y|g_c|},\qquad n\geq1.
 \label{smeq:classD_interface_ladder}
\end{equation}
The transverse level spacing is $\mathcal O(L_y^{-1/2})$, set by the
interface width $\ell_c\propto L_y^{1/2}$. The full strip has width
$w\propto L_y$. \Cref{fig:classD_droplets}(d) plots the chiral
dispersions in \cref{smeq:classD_edge_dispersion} together with the
first three massive levels in \cref{smeq:classD_interface_ladder}.

\begin{figure}[t]
 \centering
 \includegraphics[width=.8\linewidth]{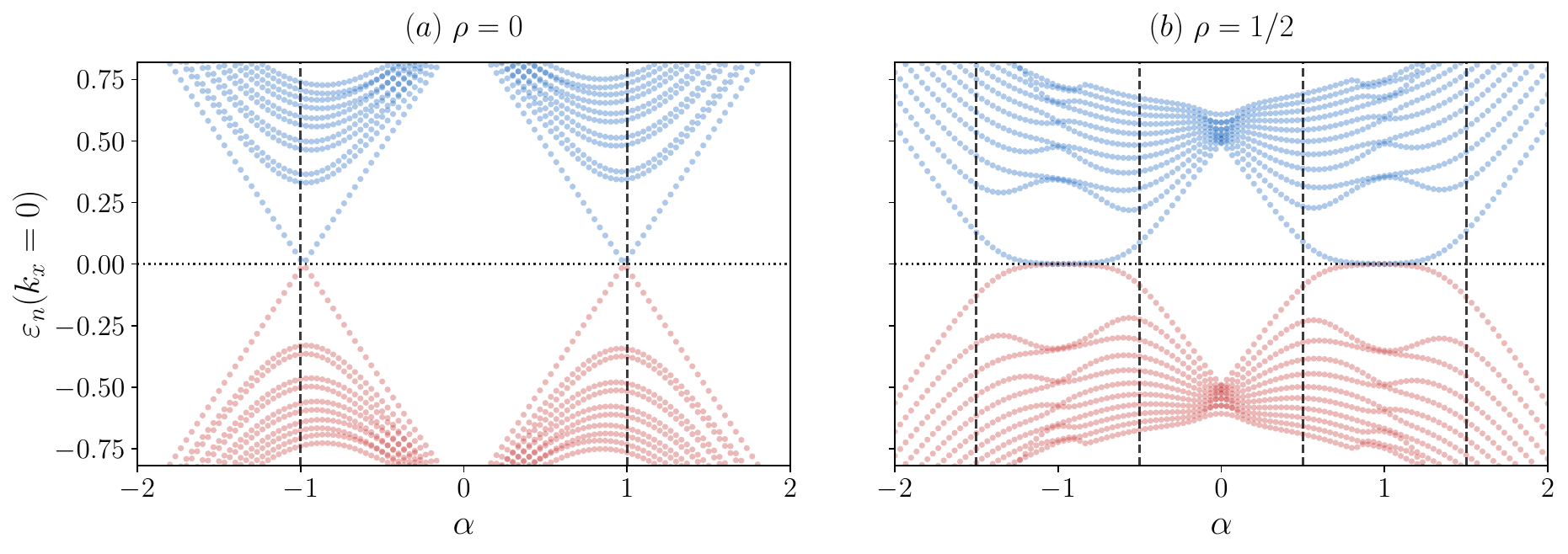}
 \caption{Numerical spectral flow across the texture
 transition. We diagonalize the lattice Hamiltonian in
 \cref{smeq:classD_H0,smeq:classD_Hrho}, with the profile in
 \cref{smeq:classD_spatial_texture}, periodic transverse boundaries,
 $k_x=0$, and $L_y=48$ two-layer cells.
 (a) At $\rho=0$, the low-energy levels approach the isolated
 trap-critical points $\alpha=\pm1$: finite-size corrections shift the
 lowest levels from their continuum values.
 (b) At $\rho=1/2$, the dashed lines delimit the TSc intervals
 $1/2<|\alpha|<3/2$ in \cref{smeq:classD_alpha_phases}. The nearly
 zero-energy branches inside these intervals arise from the two
 spatially separated interface modes. Both panels show lattice
 eigenvalues. The corresponding continuum dispersions are shown in
 \cref{fig:classD_droplets}(d).}
 \label{fig:classD_spectral_flow}
\end{figure}

At fixed parameters inside a lobe, $w/\ell_c\to\infty$ as the sample
grows, and the two interface channels become well separated. Tunneling
between them is exponentially small in $L_y$. Their longitudinal
levels have a separate spacing $\mathcal O(v_x/L_x)$. The sample thus
contains locally gapped bulk regions together with low-energy chiral
interfaces, and its full spectrum is not gapped in the thermodynamic
limit. If the sample is open along $x$, the channels join around the
ends of the strip to form a chiral perimeter mode. With periodic $x$,
they appear as two counterpropagating branches.

The nearly zero-energy branches in \cref{fig:classD_spectral_flow}(b)
are the interface states shown in \cref{fig:classD_droplets}(c,d).

\FloatBarrier
\subsubsection{The quadratic trap at a lobe phase boundary}
The separated-interface description breaks down as the strip
disappears. At a lobe phase boundary, the two zeros of $\mu_L(y)$ merge at
a point $y_*$ inside the sample. In parameter space, the path
$(m(y),\delta(y))$ is then tangent to the homogeneous critical circle
$R=\rho$, where we continue to take $\rho>0$. At each of the four
phase boundaries $\alpha_c=\pm(1\pm\rho)$, both $\mu_L(y_*)=0$ and
$\partial_y\mu_L(y_*)=0$. The leading spatial dependence of the mass
is therefore quadratic. With the parameter offset $t=\alpha-\alpha_c$ at fixed $\rho$ and the local angular
displacement $q=2\pi(y-y_*)/L_y$, expansion of $R^2$ gives
\begin{equation}
 R^2=\rho^2+2\rho A_ct+|\alpha_c|q^2
 +\mathcal O(t^2,tq^2,q^4),\qquad
 A_c=\operatorname{sgn}(\alpha_c)
       \operatorname{sgn}(|\alpha_c|-1).
\end{equation}
Consequently,
\begin{equation}
 \mu_L=R-\rho=A_ct+
 \frac{|\alpha_c|}{2\rho}
 \left[\frac{2\pi(y-y_*)}{L_y}\right]^2+\cdots,
 \qquad |A_c|=1.
 \label{smeq:classD_quadratic_mass}
\end{equation}
The sign $A_c$ specifies which direction of the parameter offset $t$ enters
the TSc region. At the outer phase boundary of the $\alpha>0$ lobe, $\alpha_c=1+\rho$ and
$A_c=1$, recovering the mass written in the main text.

Only one Majorana becomes critical on the circle. The other retains
a mass $2\rho$. We extract the critical sector directly from the lattice
Bloch Hamiltonian $h=h_0+h_\rho$ in
\cref{smeq:classD_first_quantized,smeq:classD_rho_Bloch}.
First freeze $m$ and $\delta$ at each position. At zero momentum,
\begin{equation*}
 h(\vec{0};y)=m(y)\Gamma_3+\delta(y)\Gamma_4+\rho\Gamma_D.
\end{equation*}
Here $\Gamma_D$ is defined in \cref{smeq:classD_continuum_Hamiltonian}.
With $\Gamma_R=(m\Gamma_3+\delta\Gamma_4)/R$, the projector
$P_L=(1-\Gamma_D\Gamma_R)/2$ selects the two-component sector
with signed mass $\mu_L=R-\rho$. The complementary sector has
mass $R+\rho$ and remains gapped near the transition.

To perform the projection explicitly, write
$m=R\cos\varphi$ and $\delta=R\sin\varphi$, where $\varphi(y)$ is
the angle of the local mass vector in the $(m,\delta)$ parameter plane.
A real orthonormal basis for the light sector is
\begin{equation*}
 B(y)=U(y)\frac{1}{\sqrt2}
 \begin{pmatrix}\sigma_z\\-\sigma_z\end{pmatrix},
 \qquad U(y)=e^{-\varphi(y)\Gamma_3\Gamma_4/2},
 \qquad B^{\mathsf T}B=\mathbb1,\quad BB^{\mathsf T}=P_L.
\end{equation*}
The two columns of $B$ span this sector. Expanding the full lattice
Hamiltonian to first order in momentum and projecting gives
\begin{equation}
 \begin{aligned}
 h_{\mathrm{light}}(\vec{k};y)&=B^{\mathsf T}h(\vec{k};y)B
 =v_x(y)k_x\sigma_x+v_y(y)k_y\sigma_z
 +[R(y)-\rho]\sigma_y+\mathcal O(k^2),\\
 v_x(y)&=\frac{1-m(y)}2,\qquad
 v_y(y)=\frac{1-\delta(y)}2+\frac{\rho\delta(y)}{2R(y)}.
 \end{aligned}
 \label{smeq:classD_lattice_projection}
\end{equation}
The second term in $v_y$ comes from the momentum dependence of
$h_\rho$. At a critical point $(m_*,\delta_*)$ with $R=\rho$,
$v_y=1/2$ and $v_x=(1-m_*)/2$, in agreement with the exact lattice
spectrum. We use the constant rescaling of $x$ specified above to set
both critical velocities to $1/2$.

For the smooth spatial texture, replacing momentum by
$-i\vec{\nabla}$ gives the differential operator acting on the
two-component Majorana field $\chi$, with $\Psi\simeq B(y)\chi$.
Derivatives of the local basis and velocities generate additional
gradient terms. At fixed nonzero $\rho$, these are of order $L_y^{-1}$
or smaller and are subleading to the $L_y^{-2/3}$ trap-energy scale.
Terms quadratic in momentum are also subleading.
The full local eigenvalue $R-\rho=\sqrt{m^2+\delta^2}-\rho$
includes the contribution quadratic in $\delta$ to the trap curvature.

Substituting the mass profile in \cref{smeq:classD_quadratic_mass}
into the projected Hamiltonian gives
\begin{align}
 \Ham_{\mathrm{trap}}^{(\rho\ne0)}&\approx\frac12\int d^2r\,
 \chi^\dagger h_{\mathrm{light}}\chi,\nonumber\\
 h_{\mathrm{light}}&=-\frac{i}{2}(\sigma_x\partial_x+\sigma_z\partial_y)
 +\left[A_ct+c\frac{(y-y_*)^2}{L_y^2}\right]\sigma_y+\cdots,
 \qquad c=\frac{2\pi^2|\alpha_c|}{\rho}.
 \label{smeq:classD_light_Majorana}
\end{align}
Here $\chi=[\chi^\dagger]^T$ is a two-component Majorana field, and
$c$ is the curvature coefficient of the mass profile. For negative
$\rho$, the same lengths and gaps follow with $\rho$ replaced by
$|\rho|$. At $t=0$, the relativistic dispersion has $z=1$ and
the quadratic spatial mass has $p=2$. Applying
\cref{smeq:trap_scaling} gives
\begin{equation}
 \vartheta=\frac{p}{p+z}=\frac23,\qquad
 \ell_{\mathrm{tan}}\propto L_y^{2/3},\qquad
 \Delta_{\mathrm{trap}}\propto L_y^{-2/3},\qquad
 \mathcal A_{\mathrm{crit}}\sim L_x\ell_{\mathrm{tan}}.
 \label{smeq:classD_tangent_exponent}
\end{equation}
The critical region extends along $x$. Inside the lobe, the TSc strip
instead occupies the extensive area $L_xw$
~\cite{CampostriniVicari2009PRL_TSS,CampostriniVicari2010PRA_QTSS}.

At fixed $|\rho|$, define the detuning $\tau=|\rho|-R_{\min}$, with
$R_{\min}$ given in \cref{smeq:classD_R_texture}.
It is positive inside a lobe, and the minimum light mass is $-\tau$.
Its RG eigenvalue is
$y_M=1$, so trap scaling gives the crossover window
$\tau\sim L_y^{-2/3}$. Within this window, the strip width and interface
width are both of order $L_y^{2/3}$, and the two interfaces merge into
the quadratic trap. The single-Majorana projection
requires the other sector's gap to exceed the trap energy scale, or
$\rho\sqrt{L_y}\gg1$. When $\rho\sqrt{L_y}\lesssim1$, the two-Majorana
linear-trap description is recovered. The limits $L_y\to\infty$ and
$\rho\to0$ therefore do not commute.

\subsection{Unwinding the texture and defect transitions}
The texture studied so far traverses a full parameter circle across the
sample. As in the Rice--Mele example, we can unwind it by reducing the
angle traversed across the sample. Introduce
$0\leq\beta\leq1$, so that the retained fraction of the circle is
$1-\beta$:
\begin{equation}
 \zeta_\beta(y)=\frac{2\pi(1-\beta)y}{L_y},\qquad
 m(y)=\alpha+\sin\zeta_\beta(y),\qquad
 \delta(y)=\cos\zeta_\beta(y),\qquad 0\leq\beta\leq1.
 \label{smeq:classD_texture}
\end{equation}
At $\beta=0$ the sample traverses the full circle in
\cref{smeq:classD_spatial_texture}. At $\beta=1$ the couplings are spatially
uniform.
For $0<\beta<1$, the parameters at the two endpoints differ. In an open
sample, the profile remains smooth up to the edge. Joining the
endpoints to make the sample periodic creates a local seam where the
parameters jump. 

\begin{figure}[t]
 \centering
 \begin{minipage}[t]{.66\linewidth}
  \vspace{0pt}\centering
  \includegraphics[width=\linewidth]{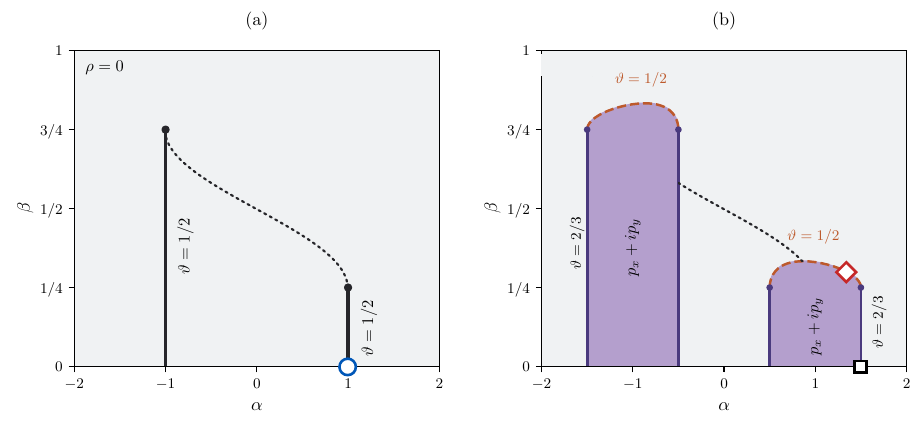}
 \end{minipage}\hfill
 \begin{minipage}[t]{.32\linewidth}
  \vspace{0pt}\centering
  \includegraphics[width=\linewidth]{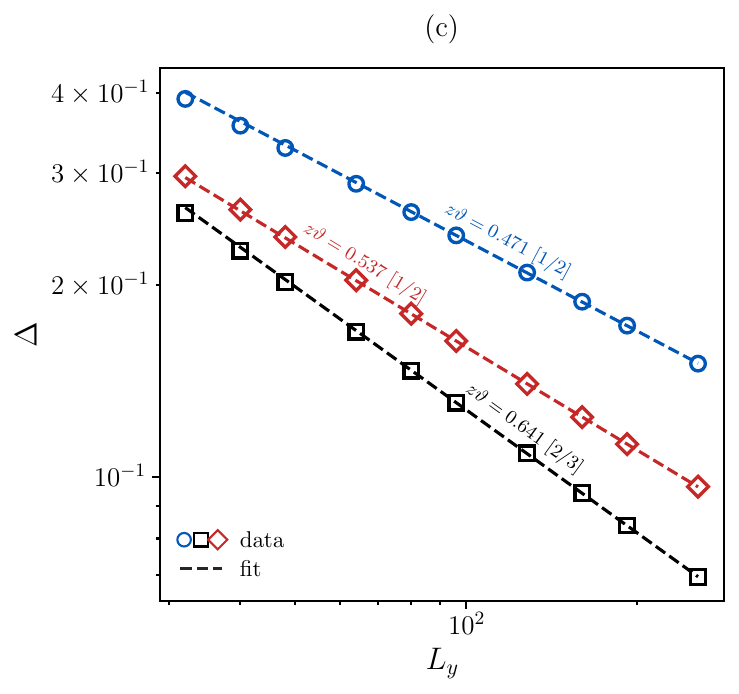}
 \end{minipage}
 \caption{Unwinding the class D texture in
 \cref{smeq:classD_texture} and testing its trap-energy scaling.
 (a) At $\rho=0$, the bulk trap-critical lines end when the position
 where the local gap vanishes reaches $y=L_y$. (b) At $\rho=1/2$, the
 TSc lobes close along the curved arcs in \cref{smeq:classD_cap_curves}.
 Along the straight phase boundaries, the two strip interfaces merge inside the
 sample and the critical mass profile is quadratic. Along the arcs,
 the remaining interface reaches the open edge, where the mass
 profile is linear. The canonical
 defect-transition locus in \cref{smeq:classD_defect_locus} is
 shown by the dotted curves. For $\rho=1/2$, the dotted curve runs from
 $(\alpha,\beta)=(\sqrt3/2,1/3)$ to $(-1/2,7/12)$. The colored markers
 locate the numerical calculations in (c):
 the blue circle is $(\alpha,\beta;\rho)=(1,0;0)$, the black square is
 $(3/2,0;1/2)$, and the red diamond is $(1.344,0.30;1/2)$.
 (c) Lattice-diagonalization data from
 \cref{smeq:classD_H0,smeq:classD_Hrho} at $k_x=0$, for
 $32\leq L_y\leq256$ two-layer cells. The circle and square use
 periodic transverse boundaries. The diamond uses an open boundary
 at which the strip disappears. For the open strip, the numerical profile
 samples $\zeta_j=2\pi(1-\beta)j/(L_y-1)$, $j=0,\ldots,L_y-1$,
 with both layers in each cell sharing $m_j=\alpha+\sin\zeta_j$
 and $\delta_j=\cos\zeta_j$. Periodic strips use $L_y$ in the denominator.
 We extract adjacent low-energy
 transverse level spacings, excluding the smaller finite-lattice shift
 of the nominal zero mode at $\rho=0$, and fit $\Delta=A L_y^b$ over the six
 largest sizes, $80\leq L_y\leq256$. The dashed curves give these
 fits: for the
 $\rho=0$ trap, straight phase boundary, and curved phase boundary, respectively,
 $(A,b)=(2.05,-0.471),(2.44,-0.641)$, and $(1.90,-0.537)$. The labels on
 the fits give the extracted $z\vartheta=-b$, with the theoretical values
 in square brackets. }
 \label{fig:classD_unwinding}
\end{figure}

At $\rho=0$, the positions where the local homogeneous gap closes move
toward the sample edge as $\beta$ increases:
$y_*/L_y=3/[4(1-\beta)]$ for $\alpha^*=1$ and
$y_*/L_y=1/[4(1-\beta)]$ for $\alpha^*=-1$.
They reach $y=L_y$ at $\beta=1/4$ and $3/4$, respectively. Thus the
bulk trap-critical lines terminate at $(\alpha,\beta)=(1,1/4)$ and
$(-1,3/4)$. Along either line, the mass varies linearly about the
trapping position and $\vartheta=1/2$. Their termination realizes the
unnecessary criticality discussed in the main text
~\cite{AP_UC_Multiversality,AP_UC_Pump} and shown in
\cref{fig:classD_unwinding}(a).

For $0<\rho<1$, the transition depends on whether the strip disappears
inside the sample or at its edge. As long as the path contains
its point of closest approach to $m=\delta=0$, the minimum of $R(y)$
occurs in the interior. The condition $R_{\min}=\rho$ then fixes the
vertical phase boundaries at constant $\alpha$. The TSc regions between these phase
boundaries exist in
\begin{align}
 1-\rho<\alpha<1+\rho,&\qquad 0\leq\beta<1/4,
 \label{smeq:classD_positive_lobe_vertical}\\
 -1-\rho<\alpha<-1+\rho,&\qquad 0\leq\beta<3/4.
 \label{smeq:classD_negative_lobe_vertical}
\end{align}
The two interfaces merge at these vertical phase boundaries, giving the
quadratic trap and $\vartheta=2/3$ derived above. Once the point of
closest approach has passed beyond $y=L_y$, the minimum lies at the
sample edge. The strip then disappears when this endpoint reaches
the homogeneous critical circle. The condition $R(L_y)=\rho$ gives
the curved phase boundaries, or caps, that close the lobes:
\begin{equation}
 \rho^2=\alpha^2+1-2\alpha\sin(2\pi\beta),\qquad
 \alpha_\pm(\beta)=\sin(2\pi\beta)
 \pm\sqrt{\rho^2-\cos^2(2\pi\beta)}.
 \label{smeq:classD_cap_curves}
\end{equation}
The curved phase boundary of the $\alpha>0$ lobe exists for
$1/4<\beta<1/4+\arcsin\rho/(2\pi)$ and closes at
$\alpha=\sqrt{1-\rho^2}$. That of the $\alpha<0$ lobe exists for
$3/4<\beta<3/4+\arcsin\rho/(2\pi)$ and closes at
$\alpha=-\sqrt{1-\rho^2}$. These straight and curved phase boundaries enclose the TSc
regions in \cref{fig:classD_unwinding}(b). Their locations are unchanged
under $\rho\to-\rho$.

The strip width can be followed throughout this deformation. Its
original angular interval is $[\theta_*-\phi_0,\theta_*+\phi_0]$, with
$\theta_*$ and $\phi_0$ defined in \cref{smeq:classD_droplet_angle}.
Only the part within $0\leq\zeta_\beta\leq2\pi(1-\beta)$ remains in
the sample. Converting this angular interval into a length gives
\begin{equation}
 w_\beta=\frac{L_y}{2\pi(1-\beta)}
 \left[\min\{\theta_*+\phi_0,\,2\pi(1-\beta)\}
       -(\theta_*-\phi_0)\right]_+,
 \qquad [z]_+=\max(z,0).
 \label{smeq:classD_shortened_width}
\end{equation}
This formula applies when $\phi_0$ in \cref{smeq:classD_droplet_angle} is real and $\beta<1$. Otherwise
the width is zero. Before either strip interface reaches the sample edge,
$w_\beta=w/(1-\beta)$, where $w$ is the width at $\beta=0$ in
\cref{smeq:classD_droplet_width}. For interfaces still inside the sample,
the finite-$\beta$ results follow by replacing $L_y$ with
$L_y/(1-\beta)$ in \cref{smeq:classD_interface_gradient} and the
definition of $\theta_c$ below it. The same replacement in the quadratic
mass profile of \cref{smeq:classD_quadratic_mass} gives
$\ell_{\mathrm{tan}}\propto[L_y/(1-\beta)]^{2/3}$ along the straight
phase boundaries, leaving the trap exponent unchanged.

Along a curved phase boundary, the last TSc interface reaches the open edge
with a nonzero mass gradient. To describe this transition, measure
the distance from the open edge by $s=L_y-y\geq0$, and write
$r_b=R(L_y)-\rho$ for the light mass at the endpoint. This detuning
is positive on the locally trivial side, vanishes at the transition,
and is negative when the endpoint lies inside the TSc strip. Expanding
$\mu_L=R-\rho$ near $s=0$ gives a linear mass on the half-space
$s\geq0$ and the corresponding Hamiltonian:
\begin{equation}
 \begin{aligned}
 \mu_L(L_y-s)&=r_b+|g_L|s+\mathcal O(s^2/L_y^2),\\
 h_{\mathrm{edge}}&=-\frac{i}{2}
 (\sigma_x\partial_x-\sigma_z\partial_s)
 +[r_b+|g_L|s]\sigma_y+\cdots,\\[-2pt]
 |g_L|&=|R'(L_y)|=\frac{2\pi(1-\beta)}{L_y}
 \frac{|\alpha\cos(2\pi\beta)|}{\rho} .
 \end{aligned}
 \label{smeq:classD_cap_theory}
\end{equation}
The gradient is evaluated on the curved phase boundary, where $r_b=0$ and
$R(L_y)=\rho$. Its sign in the expansion expresses the increase of
the mass on moving into the locally trivial interior.

At the curved phase boundary, $r_b=0$. The Majorana dispersion is
relativistic, $z=1$, and the mass vanishes linearly with the spatial
coordinate $s$, so $p=1$. Equation~\eqref{smeq:trap_scaling} therefore gives
\begin{equation*}
 \vartheta=\frac{p}{p+z}=\frac12,\qquad
 \ell\sim L_y^{1/2},\qquad
 \Delta_\perp\sim L_y^{-1/2},\qquad
 \mathcal A_{\mathrm{crit}}\sim L_xL_y^{1/2}.
\end{equation*}
The open edge fixes the Majorana boundary condition
~\cite{BerryMondragon1987}: at a generic point away from the junctions,
this affects numerical coefficients but not these scaling powers.

Along a straight phase boundary, the parameter path touches the homogeneous
critical circle at an interior position with zero mass gradient.
The leading spatial profile is quadratic, $p=2$, and $z=1$.
Equation~\eqref{smeq:trap_scaling} gives $\vartheta=2/3$, $\ell\sim L_y^{2/3}$, transverse gap
$L_y^{-2/3}$, and critical area $L_xL_y^{2/3}$.  Thus the same two phases
meet through two inequivalent universal theories, providing the class D
realization of trap-scaling multiversality~\cite{AP_UC_Multiversality}.
At the marked junctions between the straight and curved phase boundaries in
\cref{fig:classD_unwinding}(b), the interior tangency point has just
reached $y=L_y$. There $R'(L_y)=g_L=0$, so the quadratic term becomes
leading and describes the crossover between the two scaling regimes
at the sample edge.

The numerical spectra in \cref{fig:classD_unwinding}(c)
test these three trap laws at the circle, square, and diamond marked
in panels (a,b). The predicted transverse spacings scale as
$L_y^{-1/2}$ at $\rho=0$, $L_y^{-2/3}$ on a straight phase boundary,
and $L_y^{-1/2}$ on a curved phase boundary.

\subsubsection{Kitaev-chain transition at the edge}

The suspension pairs neighboring Kitaev chains into a trivial state.
An open edge can leave one chain unpaired. Unwinding the texture moves
the trapped chain toward the edge, where it can undergo its own
topological transition. For the open termination used here, its mass
is controlled by the endpoint parameters
\begin{equation}
 m_L=\alpha-\sin(2\pi\beta),\qquad
 \delta_L=\cos(2\pi\beta).
 \label{smeq:classD_endpoint_parameters}
\end{equation}
When $\delta_L<0$, the layer pairing leaves an edge chain along $x$.
Near its transition, the leading low-energy spectrum has the
Kitaev-chain form, up to finite renormalizations,
\begin{equation}
 E_\pm(k_x)\simeq\pm\sqrt{v_{\rm e}^2k_x^2+m_L^2},
\end{equation}
where $v_{\rm e}$ is the velocity along the edge. The gap closes at
$m_L=0$, giving a massless one-dimensional Majorana fermion in the
Ising universality class, with $c=1/2$ and $z=1$
~\cite{KitaevMajoranaChain,Cardy_ScalingRG_StatPhys_1996}.
The transition therefore occurs along the phase-diagram locus
\begin{equation}
 \alpha_{\rm def}(\beta)=\sin(2\pi\beta).
 \label{smeq:classD_defect_locus}
\end{equation}

At $\rho=0$, this line extends over $1/4<\beta<3/4$ and joins the
terminal points of the bulk trap-critical lines. For $0<|\rho|<1$,
the segment outside the TSc lobes separates the two gapped texture
sectors: it joins the curved phase boundary of the $\alpha>0$ lobe
to the inner phase boundary of the $\alpha<0$ lobe, as shown by the
dotted curve in \cref{fig:classD_unwinding}(b).
The critical mode propagates along $x$ and remains localized near
$y=L_y$. Away from the junctions with bulk criticality, its transverse
width stays finite and its nonzero excitation spacing scales as
$L_x^{-1}$. Changing the edge couplings or joining the sample into
a periodic seam can shift the transition locus.

\FloatBarrier
\subsection{Perturbative stability and the distinct spatial scales}
The effect of $\rho$ illustrates the stability question: an allowed
local perturbation splits the two Majorana species at the $\rho=0$
trap and changes its phase diagram. At a lobe phase boundary with
finite $\rho$, however, only one two-component Majorana remains light.
At zero momentum, the BdG particle--hole constraint in
\cref{smeq:classD_PHS} requires its Hamiltonian to be imaginary and
antisymmetric. The only Hermitian $2\times2$ matrix with these
properties is proportional to $\sigma_y$. Thus the mass already
included in \cref{smeq:classD_light_Majorana} exhausts the allowed
relevant quadratic perturbations. Once a fixed $0<|\rho|<1$ is included,
further weak local perturbations leave the phase diagrams in
\cref{fig:classD_homogeneous,fig:classD_unwinding} qualitatively unchanged.
They shift the phase boundaries, the position where the strip first
appears, and the velocities. A quadratic minimum with nonzero
curvature persists under these changes, preserving $\vartheta=2/3$.

Weak short-range interactions can be assessed by power counting at
the free relativistic fixed point~\cite{Cardy_ScalingRG_StatPhys_1996}.
In $2+1$ spacetime dimensions, a
Majorana field has scaling dimension one. A four-fermion operator
therefore has dimension four, and its coupling has RG eigenvalue
$3-4=-1$. Such interactions are irrelevant. When a contact term
vanishes for a single species, a nonzero interaction requires
derivatives and is still less relevant. The bulk trap exponents
therefore persist in an open neighborhood of the free fixed points
under weak short-range interactions.

\Cref{tab:classD_scales} summarizes the spatial and energy
scales of the different regimes.

\begin{table}[H]
 \centering
 
 \begin{tabular}{llll}
 Region & Transverse size & Area & Transverse energy scale\\\hline
 $\rho=0$ bulk trap & $L_y^{1/2}$ & $L_xL_y^{1/2}$ & $L_y^{-1/2}$\\
 TSc strip, inside a lobe & $w\propto L_y$ & $L_xL_y$ & locally gapped bulk\\
 A simple strip interface & $L_y^{1/2}$ & $L_xL_y^{1/2}$ & $L_y^{-1/2}$\\
 Straight phase boundary (interior tangency) & $L_y^{2/3}$ & $L_xL_y^{2/3}$ & $L_y^{-2/3}$\\
 Curved phase boundary (open endpoint) & $L_y^{1/2}$ & $L_xL_y^{1/2}$ & $L_y^{-1/2}$
 \end{tabular}
 \caption{Distinct spatial scales at fixed parameters as $L_y\to\infty$.
 The energy entries for a linear trap or interface refer to the massive
 transverse levels. The lowest branch propagates along $x$. The full
 TSc strip has extensive area, while the region supporting an interface
 mode and the critical region at a lobe phase boundary occupy vanishing
 fractions of the sample.}
 \label{tab:classD_scales}
\end{table}

\FloatBarrier

\section{Classifying diabolical textures and the \texorpdfstring{$\Omega$}{Omega} spectrum}
\label{app:omega_classification}

The classification assumes slowly varying, locally gapped textures
and the stated boundary conditions. It does not classify arbitrary
inhomogeneous Hamiltonians or determine the critical theory when a
texture is removed. The trap-scaling theories in the main text require
the microscopic analyses given in the preceding sections.

\subsection{From ordered-media textures to diabolical textures}
Let us revisit the analogy to ordered media that was briefly discussed in the end matter. 
Consider first an easy-plane magnet. Its local order parameter is an
angle, so the allowed values form a circle. A vortex is detected by the
winding of this angle around a closed path in real space. The order
parameter is well defined everywhere on the path, but a nonzero winding
cannot be extended to the enclosed disk while remaining on the
order-parameter circle. The core is where this local description fails.
Similarly, the unit-vector order parameter of an isotropic magnet takes
values on a two-sphere, and a hedgehog is detected by the way that vector
covers the sphere on a surface surrounding the core
~\cite{Mermin_RevModPhys.51.591,Michel_OrderedMedia_RevModPhys.52.617}. In each case, the texture assigns an allowed order-parameter value to
every point in a spatial region. If a symmetry group $G$ is spontaneously
broken to a subgroup $H$, order parameters take values in  $\sC=G/H$. On a $d$-dimensional
spatial domain $\cM_d$, the texture is therefore a continuous map
\begin{equation}
 s:\cM_d\longrightarrow\sC.
 \label{smeq:ordered_texture_map}
\end{equation}
Two configurations have the same topology if one can be smoothly
deformed into the other while remaining in $\sC$ and respecting the
boundary conditions. Such a deformation is called a homotopy.
The brackets in $[s]\in[\cM_d,\sC]$ identify maps related in this way.
Restricting $s$ to a loop or sphere surrounding a core gives the vortex
or hedgehog charge. A nonsingular example is a two-dimensional magnetic
skyrmion. Fixing the spin direction along the sample boundary allows
that boundary to be identified with a single point, turning the domain
into a sphere. The texture is then a map $S^2\to S^2$, from the
compactified sample to the sphere of spin directions
~\cite{Mermin_RevModPhys.51.591,ang2018classificationtopologicaldefectstextures,ang2018highercategoricalgroupsclassification}.

For a diabolical texture, the local information comes from the
Hamiltonian. At each position, we freeze the slowly varying couplings
to obtain a homogeneous, gapped Hamiltonian. There is no spontaneously
chosen order parameter. We fix the onsite symmetry $G$ and whether the
microscopic degrees of freedom are bosonic or fermionic: in the latter
case, $G$ includes fermion parity.

The relevant space consists of \emph{invertible} Hamiltonians. To define
invertibility, first recall that stacking places two independent
systems in parallel and treats them as one system. A gapped Hamiltonian
is invertible if it has a stacking partner such that the combined
system can be smoothly deformed to a product state without closing the
bulk gap or breaking $G$. Hamiltonians connected by gapped,
symmetry-preserving deformations belong to the same invertible phase
~\cite{kitaev1,kitaev2,Gaiotto_2019,kubota2025stablehomotopytheoryinvertible}.
Such a phase may have protected boundary
physics, as in a symmetry-protected topological or integer quantum Hall
phase, but has no intrinsic topological order in the bulk
~\cite{kitaevclassification,Xiong_2018,Gaiotto_2019}.
Let $\sI_d^G$ denote the space of local, gapped, invertible Hamiltonians
in $d$ spatial dimensions. We allow decoupled orbitals or spins in fixed,
symmetry-trivial product states to be added. This is the
\emph{stabilized} space: the classification does not depend on how many
such inert degrees of freedom are included
~\cite{kitaev1,kubota2025stablehomotopytheoryinvertible}.
For examples of stacking inverses, a Chern insulator with Chern number
$C=+1$ cancels one with $C=-1$, while two Kitaev chains can be
trivialized because the one-dimensional class D classification is
$\bZ_2$
~\cite{KitaevMajoranaChain,kitaevclassification}.
The texture is the map
\begin{equation}
 \varrho:\cM_d\longrightarrow\sI_d^G,\qquad
 x\longmapsto H_{\rm frozen}(x).
 \label{smeq:Hamiltonian_texture_map}
\end{equation}
The compactification or boundary conditions of the spatial domain
$\cM_d$ must be specified, just as for an ordered medium
~\cite{Mermin_RevModPhys.51.591,Michel_OrderedMedia_RevModPhys.52.617}.
The two constructions can be compared directly:
\begin{center}
\setlength{\tabcolsep}{8pt}
\renewcommand{\arraystretch}{1.2}
\begin{tabular}{c|c|c}
 & local datum & allowed-value space\\ \hline
 ordered medium & $s(x)$ & $\sC=G/H$\\
 diabolical texture & $H_{\rm frozen}(x)$ & $\sI_d^G$.
\end{tabular}
\end{center}
An allowed deformation keeps the order parameter in $\sC$, or the
frozen Hamiltonian gapped and symmetry preserving in $\sI_d^G$.
A diabolical point is excluded from the latter space because the gap
closes there. A nontrivial parameter loop around it therefore plays the
same role as the order-parameter loop around a vortex core: shrinking
the loop requires leaving the allowed space. For Hamiltonians, this
means closing the gap or relaxing the specified constraints. The
diabolical point lies in the Hamiltonian phase diagram. Every frozen
Hamiltonian along the spatial loop can remain gapped, so the texture
need not contain a gapless point in real space.

For an ordered medium, the broken symmetry determines the space
$\sC$. The spaces $\sI_d^G$ have an additional structure: they are
conjecturally related across spatial dimensions. This relation
associates a winding of Hamiltonians with a lower-dimensional quantum
phase, as seen in the charge bound by the Rice--Mele texture and the
Kitaev chain carried by the class D texture. It also makes the
Hamiltonian-texture classification more tractable than a general
classification of maps into an order-parameter space.

To state this relation, we need to distinguish a space of Hamiltonians
from the set of phases it contains. A connected component of $\sI_d^G$
consists of Hamiltonians joined by gapped, symmetry-preserving paths,
and hence describes one phase. Writing $\pi_0$ for the set of these
components, the group of invertible phases is
\begin{equation}
 \mathcal I_d^G=\pi_0(\sI_d^G).
 \label{smeq:omega_phase_group}
\end{equation}
Stacking defines addition in $\mathcal I_d^G$, the stacking partner
being the inverse, and a reference product state is the identity.
Thus $\mathcal I_d^G$ labels the phases,
while $\sI_d^G$ also retains the paths and families within each phase
~\cite{Gaiotto_2019,Xiong_2018}.

\subsection{The $\Omega$-spectrum descent}

Choose a reference product-state Hamiltonian $H_{\rm ref}$ and vary the
couplings around a gapped cycle $H(\lambda)$, with
$0\leq\lambda\leq1$ and $H(0)=H(1)=H_{\rm ref}$. We can regard the
whole cycle as one point in a new space, the \emph{based loop space}
$\Omega\sI_d^G$. Here ``based'' means that the cycle starts and ends at
the chosen reference, which remains fixed during a deformation. Two
cycles belong to the same connected component of this space if they
can be deformed into one another without closing the gap or breaking
$G$. Kitaev's $\Omega$-spectrum conjecture relates this space of cycles
to Hamiltonians in one lower spatial dimension:
\begin{equation}
 \Omega\sI_d^G\simeq\sI_{d-1}^G.
 \label{smeq:omega_loop_space}
\end{equation}
 The equivalence in \cref{smeq:omega_loop_space} matches the connected components of the two spaces,
as well as their homotopy groups, which classify spherical families
within them. In mathematical terms, it is a weak homotopy equivalence.
A sequence of spaces equipped with compatible equivalences of this
form is called an $\Omega$ spectrum~\cite{kitaev1,kitaev2,kitaev3,kubota2025stablehomotopytheoryinvertible}. For the physical motivation behind Kitaev's $\Omega$-spectrum conjecture
and detailed treatments of its homotopy-theoretic and
generalized-cohomology formulation, see
Refs.~\cite{kitaev1,kitaev2,kitaev3,kubota2025stablehomotopytheoryinvertible,Xiong_2018,Gaiotto_2019}.
For general topological background, see Ref.~\cite{Hatcher2002}.

We can construct such a cycle by stacking lower-dimensional systems.
Let $\mathcal P$ be a $(d-1)$-dimensional invertible phase and
$\overline{\mathcal P}$ its stacking inverse. Arrange alternating layers
of the two along a new spatial direction $x_d$. Each neighboring pair
can be coupled and deformed to a product state. There are two choices
of pairing: layers $(2j,2j+1)$ or $(2j-1,2j)$, where $j$ is an integer.
Starting from the first paired product state, undo the pairing to reach
the decoupled stack, then couple the layers in the second pairing.
Both endpoints can be chosen to be $H_{\rm ref}$, so this gapped path
forms a loop of $d$-dimensional Hamiltonians. This is the
\emph{suspension construction}
~\cite{Gaiotto_2019,Wenetal_topologicalfamilies,AP_ThoulessPhasediagrams}.

At an open termination, shifting the pairing leaves one
unpaired $\mathcal P$ layer, even though the bulk pairs are trivial.
The same lower-dimensional phase appears when the loop parameter varies
slowly along a spatial direction: the resulting $d$-dimensional texture
carries a $(d-1)$-dimensional invertible phase. Thus the connected
components of the loop space, or equivalently the loop classes
$\pi_1(\sI_d^G)$, are labeled by lower-dimensional phases:
\begin{equation}
 \pi_1(\sI_d^G)\cong\pi_0(\Omega\sI_d^G)
 \cong\mathcal I_{d-1}^G.
 \label{smeq:omega_loop_phase}
\end{equation}
The examples in the main text realize \cref{smeq:omega_loop_phase}:
\begin{center}
\setlength{\tabcolsep}{8pt}
\renewcommand{\arraystretch}{1.16}
\begin{tabular}{ccl}
 Rice--Mele loop in $d=1$ & $\longrightarrow$ & $\mathrm U(1)$ charge in $d=0$,\\
 class D loop in $d=2$ & $\longrightarrow$ & Kitaev phase in $d=1$.
\end{tabular}
\end{center}

For a family with more parameters, we repeat the loop construction
underlying \cref{smeq:omega_loop_phase}.
For any space $Y$, the group $\pi_k(Y)$ consists of maps from a
$k$-sphere into $Y$, up to deformations that fix a chosen point.
These can also be viewed as connected components of the $k$-fold loop
space: $\pi_k(Y)=\pi_0(\Omega^kY)$~\cite{Hatcher2002}.
Each use of \cref{smeq:omega_loop_space} lowers the spatial dimension
by one, giving
\begin{equation}
 \pi_k(\sI_d^G)
 \cong\pi_0(\Omega^k\sI_d^G)
 \cong\pi_0(\sI_{d-k}^G)
 =\mathcal I_{d-k}^G,\qquad 1\leq k\leq d.
 \label{smeq:omega_descent}
\end{equation}
When the $k$ parameters are embedded as spatial coordinates, the family
can carry an invertible phase in the remaining $d-k$ directions.
The \emph{background phase} is the ordinary $d$-dimensional phase of
the frozen Hamiltonian at a chosen reference position. Taking that
Hamiltonian everywhere in the sample gives a uniform system in this phase.

\subsection{Spherical-cycle textures}

In an ordered medium, we can inspect a texture by following the order
parameter along a closed path or over a spherical surface. Around a
vortex, the order-parameter angle winds along the path. Around a
hedgehog, the spin direction covers the sphere
~\cite{Mermin_RevModPhys.51.591,Michel_OrderedMedia_RevModPhys.52.617}.
These paths and surfaces are examples of \emph{spherical probes}:
a circle is $S^1$, and a spherical surface is $S^2$.
For a diabolical texture, we follow the frozen Hamiltonian in
\cref{smeq:Hamiltonian_texture_map} along the probe.
A closed path then gives a gapped loop of Hamiltonians, such as
the Rice--Mele loop carrying a charge or the class D loop carrying a
Kitaev phase. More generally, a $k$-dimensional spherical probe gives
a family whose label is an invertible phase in $d-k$ dimensions,
by \cref{smeq:omega_descent}. On a two-dimensional torus, for example,
either coordinate loop can detect a one-dimensional phase.

To apply this result to a given texture, choose a position
$x_0\in\cM_d$ and let $\mathcal B$ be the phase of its frozen
Hamiltonian $\varrho(x_0)$. We first remove this uniform background by
stacking the whole texture with a uniform system in the inverse phase
$\overline{\mathcal B}$. The Hamiltonian at $x_0$ now lies in
$\mathcal B\oplus\overline{\mathcal B}$ and can be deformed to
$H_{\rm ref}$. This leaves the texture relative to its background
unchanged. For this spherical-cycle construction, $\varrho$ denotes
the family after background subtraction, based at $H_{\rm ref}$.

Mathematically, a spherical probe is a map $f:S^k\to\cM_d$.
The map $f$ places each
point of the sphere in the sample and sends its chosen basepoint to
$x_0$. Reading off the frozen Hamiltonian at those positions gives
\begin{equation}
 S^k\xrightarrow{\,f\,}\cM_d
 \xrightarrow{\,\varrho\,}\sI_d^G.
 \label{smeq:omega_composed_map}
\end{equation}
The composite map $\varrho\circ f$ in \cref{smeq:omega_composed_map}
is a spherical family of Hamiltonians. By \cref{smeq:omega_descent},
its deformation class gives the lower-dimensional phase label
\begin{equation}
 \varrho_k([f])=[\varrho\circ f]
 \in\pi_k(\sI_d^G)\cong\mathcal I_{d-k}^G,
 \qquad 1\leq k\leq d.
 \label{smeq:omega_induced_map}
\end{equation}
Here $[f]$ denotes the spherical probe up to deformation in the sample,
and $[\varrho\circ f]$ denotes the induced Hamiltonian family up to
gapped deformation. Combining probes adds their phase labels, so
$\varrho_k:\pi_k(\cM_d)\to\mathcal I_{d-k}^G$ is a group
homomorphism.
The target group is abelian: stacking phase labels commutes, even when
the order of traversing spatial loops is important in $\pi_1(\cM_d)$.
The $k=0$ datum simply assigns an ordinary $d$-dimensional phase to
each connected component of $\cM_d$: it is a map of sets, not a
homomorphism.

The maps $\{\varrho_k\}$ in \cref{smeq:omega_induced_map} provide
useful diagnostics, but spherical
probes need not detect all the structure of a texture on a general
domain. In addition, phase labels measured on different parts of the
sample must arise from one continuous, gapped Hamiltonian family.

\subsection{Full classification and compatibility}

The full classification retains the uniform background as well.
Returning to the original texture map $\varrho$ in
\cref{smeq:Hamiltonian_texture_map}, we allow deformations
that keep every frozen Hamiltonian gapped and preserve $G$. As in the
ordered-medium problem, the resulting classes
are denoted by brackets. The $\Omega$-spectrum structure identifies
them with a generalized-cohomology group:
\begin{equation}
 [\cM_d,\sI_d^G]\cong h_G^d(\cM_d),
 \label{smeq:omega_full_classification}
\end{equation}
with a given texture represented by $[\varrho]$. The notation
$h_G^d(\cM_d)$ denotes the group assigned to $\cM_d$ by the theory
represented by the $\Omega$ spectrum~\cite{Hatcher2002}. In our setting,
its elements identify diabolical textures whose spatial coupling
profiles can be smoothly deformed into one another while every frozen
Hamiltonian remains gapped and preserves $G$. This includes compatibility
between different regions. Addition stacks systems with their textures.
One can see why this is a \emph{generalized}
cohomology theory by considering a parameter space consisting of a
single point, denoted $\mathrm{pt}$. There is then no parameter
variation, but the Hamiltonian still has an ordinary phase. Our
grading convention is $h_G^q(\mathrm{pt})=\mathcal I_q^G$, where $q$
is the spatial dimension of that Hamiltonian. These groups are called
the coefficients of the theory. They can be nonzero in many degrees,
unlike the coefficients of ordinary cohomology~\cite{Hatcher2002,Xiong_2018}.
The subscript $G$ specifies the onsite symmetry of the
Hamiltonians~\cite{Xiong_2018,Gaiotto_2019}.

To extract the part due to spatial variation, suppose that the domain
is connected and choose a basepoint $x_0$. We can separate the uniform
background phase from the texture:
\begin{equation}
 h_G^d(\cM_d)\cong\mathcal I_d^G\oplus
 \widetilde h_G^d(\cM_d).
 \label{smeq:omega_reduced_classification}
\end{equation}
The first term in \cref{smeq:omega_reduced_classification} represents
the phase of $\varrho(x_0)$. Taking that
Hamiltonian everywhere in the sample gives the corresponding constant
map. Stacking the texture with the inverse of this uniform system
removes the background and leaves an element of the \emph{reduced}
group $\widetilde h_G^d(\cM_d)$. Thus the overall phase of matter and texture within that phase are represented separately.

Boundary conditions introduce complexities just as they do for
order-parameter textures. If the Hamiltonian is fixed to $H_{\rm ref}$
on the entire boundary $\partial\cM_d$, all boundary points have the
same image under $\varrho$. We can therefore identify them with one
point. The resulting space is written $\cM_d/\partial\cM_d$, and
the \emph{relative} classification is
\begin{equation}
 h_G^d(\cM_d,\partial\cM_d)
 \cong\widetilde h_G^d(\cM_d/\partial\cM_d).
 \label{smeq:omega_relative_classification}
\end{equation}
For example, an interval with free endpoints can be contracted to a
point, so it has no reduced texture invariant. Fixing both endpoints
to the same reference Hamiltonian turns the quotient in
\cref{smeq:omega_relative_classification} into a circle
and allows the loop invariant of \cref{smeq:omega_loop_phase}.

The suspension-loop correspondence links the cohomology classification
in \cref{smeq:omega_full_classification} to the dimension-lowering
relation in \cref{smeq:omega_descent}. Let $X$ be a parameter
space with a chosen basepoint. Add an interpolation coordinate in
$[0,1]$, then identify both ends and the line through the basepoint
with a single point. This space is the \emph{reduced suspension}
$\Sigma X$. Maps from $\Sigma X$ to
any based space $Y$ are equivalent to maps from $X$ to the based loop
space of $Y$~\cite{Hatcher2002}. Taking the Hamiltonians to have spatial dimension $n+1$
and using \cref{smeq:omega_loop_space} gives
\begin{align}
 \widetilde h_G^{n+1}(\Sigma X)
 &=[\Sigma X,\sI_{n+1}^G]_*
 \cong[X,\Omega\sI_{n+1}^G]_*
 \cong\widetilde h_G^n(X).
 \label{smeq:omega_suspension_axiom}
\end{align}
The subscript $*$ indicates that maps and their deformations preserve
the basepoint.

Successive spheres are related by suspension,
$\Sigma S^k\simeq S^{k+1}$~\cite{Hatcher2002}.
Repeating \cref{smeq:omega_suspension_axiom} gives
$\widetilde h_G^d(S^k)\cong\mathcal I_{d-k}^G$ for $1\leq k\leq d$:
spherical $k$-parameter families of $d$-dimensional Hamiltonians,
with a fixed background, are classified by invertible phases in
$d-k$ dimensions. When the parameters vary along a spherical spatial
cycle, a nontrivial family gives a diabolical texture trapping this
phase in the remaining $d-k$ directions. The map $\varrho_k$ in
\cref{smeq:omega_induced_map} assigns this trapped phase to the
spherical cycle $[f]$.

How do textures trapping lower-dimensional phases fit together across
the sample? Can their phase labels be assigned independently?
In general, the assignments must satisfy compatibility conditions.
 Physically, compatibility means that
the spatial coupling profiles join into one continuous family: the
frozen Hamiltonians agree where regions meet, remain gapped, and
preserve $G$. We can organize this question by dividing the sample
into vertices, edges, faces, and higher-dimensional cells.
The Atiyah--Hirzebruch spectral sequence tests the assignments
successively as we extend the family across these cells. It also
identifies assignments related by gapped deformations. The surviving
labels, together with their stacking relations, determine the full
texture classification in \cref{smeq:omega_full_classification}
~\cite{AtiyahHirzebruch1961,McCleary2001}.
For physics-oriented introductions and applications, see
Refs.~\cite{SongHuangFuHermele2017,ElseThorngren2019,
WangNingCheng2021,RenNingQiWangGu2024,
ShiozakiSatoGomi2022,ShiozakiXiongGomi2023}.

\subsection{Textures on a torus}

The spherical-probe labels $\varrho_k$ defined in
\cref{smeq:omega_induced_map} do not classify all diabolical textures.
For example, on a spatial two-torus $T^2$, the
classification of two-dimensional Hamiltonian textures is
\begin{equation}
 h_G^2(T^2)\cong\mathcal I_2^G\oplus
 (\mathcal I_1^G)^{\oplus2}\oplus\mathcal I_0^G.
 \label{smeq:omega_torus_example}
\end{equation}
The first term gives the uniform background phase, as in
\cref{smeq:omega_reduced_classification}. The two
$\mathcal I_1^G$ terms give the one-dimensional phases trapped by
textures along the two coordinate loops, as in
\cref{smeq:omega_loop_phase}. The additional
$\mathcal I_0^G$ term describes a trapped zero-dimensional phase that
can be nontrivial even when both loop labels and the background vanish.
The loop probes miss this contribution, and higher-dimensional
spherical probes cannot detect it since $\pi_{k\geq2}(T^2)=0$.
Thus the full texture contains information
beyond $\{\varrho_k\}$. The direct-sum decomposition in
\cref{smeq:omega_torus_example} follows from the torus geometry
~\cite{Hatcher2002}.

To specialize \cref{smeq:omega_torus_example}, take class D with only fermion parity,
$G=\bZ_2^f$. Here $\mathcal I_2^G=\bZ$, while
$\mathcal I_1^G=\mathcal I_0^G=\bZ_2$
~\cite{kitaevclassification,KitaevMajoranaChain}.
The integer counts the signed number of $p_x\pm ip_y$ TSc copies in
the uniform background. The next two $\bZ_2$ factors indicate whether
each coordinate-loop texture traps a Kitaev chain. The main-text
example realizes one of these: varying the couplings along $y$ traps
a chain along $x$ in a trivial background. Exchanging $x$ and $y$
realizes the other. The final $\bZ_2$ gives the even or odd fermion
parity of a trapped pointlike phase.

\end{document}